\documentclass[onecolumn,useAMS,usenatbib]{mn2e}
\usepackage{graphicx}
\usepackage{natbib}

\usepackage{bm}
\usepackage{amsfonts}
\usepackage{latexsym}
\usepackage[latin1]{inputenc}
\usepackage{amsmath}
\usepackage{epsfig}
\usepackage{amsbsy}

\newcommand{\n}{\mathrm n}
\newcommand{\f}{\mathrm f}
\newcommand{\ba}{\mathrm b}

\newcommand{\p}{\mathrm p}

\newcommand{\x}{\mathrm x}
\newcommand{\y}{\mathrm y}
\newcommand{\ch}{\mathrm c}

\newcommand{\veps}{\varepsilon}

\def \nn  {\nonumber}

\def \eps{\epsilon}
\def \veps{\varepsilon}
\def\jnl@style{\it}
\def\aaref@jnl#1{{\jnl@style#1}}

\def\aaref@jnl#1{{\jnl@style#1}}

\def\aj{\aaref@jnl{AJ}}                   
\def\apj{\aaref@jnl{ApJ}}                 
\def\apjl{\aaref@jnl{ApJ}}                
\def\apjs{\aaref@jnl{ApJS}}               
\def\apss{\aaref@jnl{Ap\&SS}}             
\def\aap{\aaref@jnl{A\&A}}                
\def\aapr{\aaref@jnl{A\&A~Rev.}}          
\def\aaps{\aaref@jnl{A\&AS}}              
\def\mnras{\aaref@jnl{MNRAS}}             
\def\prd{\aaref@jnl{Phys.~Rev.~D}}        
\def\pre{\aaref@jnl{Phys.~Rev.~E}}        
\def\prl{\aaref@jnl{Phys.~Rev.~Lett.}}    
\def\qjras{\aaref@jnl{QJRAS}}             
\def\skytel{\aaref@jnl{S\&T}}             
\def\ssr{\aaref@jnl{Space~Sci.~Rev.}}     
\def\zap{\aaref@jnl{ZAp}}                 
\def\nat{\aaref@jnl{Nature}}              
\def\aplett{\aaref@jnl{Astrophys.~Lett.}} 
\def\apspr{\aaref@jnl{Astrophys.~Space~Phys.~Res.}} 
\def\physrep{\aaref@jnl{Phys.~Rep.}}      
\def\physscr{\aaref@jnl{Phys.~Scr}}       

\title[Real neutron star seismology]{Towards real neutron star seismology: Accounting for elasticity and superfluidity}

\author[A. Passamonti $\&$ N. Andersson]
{A. Passamonti$^{1,2}$\thanks{E-mail:andrea.passamonti@uni-tuebingen.de} , N. Andersson$^2$
\\ \\
$^1$ Theoretical Astrophysics, University of T\"{u}bingen, Auf der
Morgenstelle 10, T\"{u}bingen 72076, Germany \\ 
$^2$ School of Mathematics, University of Southampton, Southampton SO17 1BJ, UK}

\begin{document}

\date{\today}

\pagerange{\pageref{firstpage}--\pageref{lastpage}} \pubyear{}

\maketitle

\label{firstpage}


\begin{abstract}

  We study the effects of an elastic crust on the oscillation spectrum
  of superfluid neutron stars.  Within the two fluid formalism, we
  consider Newtonian stellar models that include the relevant 
  constituents of a mature neutron stars. The core is formed by a
  mixture of superfluid neutrons and a conglomerate of charged
  particles, while the inner crust is described by a lattice of nuclei
  permeated by superfluid neutrons.  We linearise the
  Poisson and the conservation equations of nonrotating superfluid
  stars and study the effects of elasticity, entrainment and
  composition stratification on the shear and acoustic modes. In both
  the core and the crust, the entrainment is derived from 
  recent results for the nucleon effective mass.  Solving the
  perturbation equations as an eigenvalue problem, we find that the presence of
  superfluid neutrons in the crust and their large effective mass may have
  significant impact on the star's oscillation spectrum.

\end{abstract}

\begin{keywords}
methods: numerical -- stars: neutron -- stars: oscillation.
\end{keywords}

\section{Introduction} \label{sec:Int}

Neutron stars represent very complex systems, the understanding of which relies on uncertain theory combined with largely indirect observational evidence. The nature of the deep neutron star core, and the state of matter within, remains 
poorly constrained. However, it is clear that the quality of the gathered data, and our interpretation of it, is improving. As evidence of recent progress, with direct impact on the work discussed in this paper, we may consider the quasiperiodic oscillations observed in the tails of giant magnetar flares~\citep{2005ApJ...628L..53I, 2006ApJ...653..593S, 2006ApJ...637L.117W}. The interpretation of the data in terms of torsional oscillations of the neutron star's crust provides us (at least in principle) with  a handle on the associated physics. These observations have led to a renewed interest in neutron star seismology, and work addressing issues ranging from the crust equation 
of state~\citep{PhysRevLett.103.181101}, to the role of the magnetic field (and the coupling to the fluid core)~\citep{2006MNRAS.371L..74G,2011MNRAS.410.1036V, 2011MNRAS.tmp..566C,2011MNRAS.410L..37G} and even the relevance of the superfluid neutrons that permeate the nuclear lattice at densities beyond neutron 
drip~\citep{2009CQGra..26o5016S, 2009MNRAS.396..894A}.
A similar breakthrough is hinted at by the observed cooling rate of the compact remnant in Cassiopeia A. The data provides clear evidence for the presence of superfluid neutrons in the star's core (associated with additional cooling due to the Cooper pair breaking mechanism), leading to a constraint on the superfluid transition 
temperature~\citep{2011PhRvL.106h1101P, 2011MNRAS.412L.108S}. This constrains the theoretical models, which depend crucially on many-body interactions at extreme densities. Combined, the information gleaned about the crust region and 
the superfluid core provides us with a clear motivation to improve our seismology models accordingly. This is not a trivial task, but considerable progress has been made on the key aspects, realistic models would now seem to be within reach.

The work reported in this paper takes a serious step towards the modelling of real neutron star seismology. Our effort should be considered in the context of previous work, such as~\cite{1988ApJ...325..725M} and~\cite{1991ApJ...372..573S} who accounted for the crust elasticity and finite temperature effects, the modelling of superfluid dynamics by (for example)~\cite{1994ApJ...421..689L} and most recently~\cite{2009MNRAS.396..951P} and~\cite{2011MNRAS.413...47P} in Newtonian gravity 
and~\cite{2008PhRvD..78h3008L}, \cite{2009CQGra..26o5016S} and \cite{2011arXiv1105.4040K} in the framework of General Relativity, 
and finally the effort aimed at including the magnetic field (obviously a key aspect for magnetars!)~\citep{2011MNRAS.tmp..566C,2011MNRAS.410L..37G}.  
These studies provide key insights into the individual pieces of physics, but do not combine them. It is obviously desirable to do so, and the present work takes important steps in this direction. 
We consider the oscillations of a neutron star model with an elastic crust permeated by superfluid neutrons and a fluid core
with two dynamically distinct components. We develop, and implement, the relevant conditions at the interface between these two regions.  We do not, however, account for the magnetic field. Neither do we (fully) account for thermal effects or consider the problem in General Relativity. 
These are obvious shortcomings of our final model, especially since we cannot meaningfully make use of a truly realistic 
equation of state unless we work in relativistic gravity. However, our model still represents the state-of-the-art for the coupled 
multi-fluid-elastic aspects and the analysis provides valuable insights that will assist future developments in this problem area. 

\section{The two ``fluid'' model} \label{sec:2f-model}

We want to build a neutron star model that accounts for the degrees of freedom associated with the crust elasticity and the 
superfluid components, both in the crust region and in the core. It is well-known that 
this involves modelling the core  as a 
two-fluid system, with the superfluid neutron being treated as distinct from a charge-neutral conglomerate of protons and electrons.
If the latter components are strictly co-moving, and we ignore issues associated with superconductivity and the presence 
of magnetic fluxtubes~\citep*{2011MNRAS.410..805G}, then we may ignore magnetic field effects altogether. This is obviously not a
 true representation of a neutron star core, which should be magnetized, but the model is nevertheless useful as it allows us to explore 
 the relevance of the additional degree of freedom implied by superfluidity.

\subsection{The dynamics}

In the outer core of a neutron star, we expect superfluid neutrons and
superconducting protons to be present. Meanwhile, in the inner crust, at densities above
the neutron drip threshold, a fraction of neutrons are
superfluid and coexist with the lattice of heavy nuclei.  As we approach the
neutron drip density, these ``free'' neutrons disappear and all
the constituents of the crust are bound in nuclei.
A two ``fluid'' formalism is sufficient to describe these key aspect of neutron star dynamics, but we have to 
carefully consider the nature of each  
component in the crust and in the core.  In the core it is natural to distinguish the
superfluid neutrons (denoted by n in the following) from a conglomerate of charged particles (protons/electrons, 
denoted by p). The distinction in the inner crust is not quite so obvious, see the discussion by~\cite{2011arXiv1105.1244A}.
In this case, we consider the dynamics
of baryons confined in the nuclei (denoted by c)
permeated by a gas of free superfluid neutrons (represented by f).

With these definitions, the component fractions of the crust and the
core are in general discontinuous at the crust/core interface, even if
the total number of neutrons and protons is continuous. Therefore, the 
choice of junction conditions at the crust/core transition is
very important for the oscillation dynamics of the star. 

The dynamics of a two-component star are governed by mass and
momentum conservation equations for each constituent, together with the Poisson
equation for the gravitational potential~\citep{2004PhRvD..69d3001P}. The conservation equations take
the form:
\begin{equation}
\frac{\partial \rho _{\x} }{\partial t } + \nabla_{i}  \left( \rho
  _{\x}  v_{\x}^{i} \right)  = 0 \,  ,  \label{eq:mconv}
\end{equation}
\begin{equation}
(\partial_t + v_\x^j \nabla_j ) (v^\x_i+\varepsilon_\x w^{\y\x}_i) +\nabla_i (\tilde{\mu}_\x+\Phi)
+ \varepsilon_\x w^j_{\y\x} \nabla_i v^\x_j= f^\x_i/\rho_\x \ ,  
\label{Eulers}
\end{equation}
where $v_\x^i$ and $w_{\x\y}^i = v_\x^i-v_\y^i$ are, respectively, the
constituent and the relative velocities, while
$\tilde{\mu}_\x=\mu_\x/m_\x$ represents the chemical potential 
(we will assume that the neutron and proton masses are
equal,  i.e., we take $m=m_\p=m_\n$).  
The mass densities are defined by $\rho_\x=m
n_\x$, where $n_{\x}$ is the constituent number density. The
gravitational potential $\Phi$ obeys the Poisson equation,
\begin{equation}
\nabla^2 \Phi =  4 \pi G  \rho  \,  ,  \label{eq:Pois}  
\end{equation}
where $\rho = \sum_{\x} \rho_\x $ is the total mass density. In the interest of economy we will,
whenever we write down an equation that describes both the dynamics in the core and the crust,
imply that the two crust components have x either ``c'' or ``f''.  In the
core, these constituent indices should be replaced by ``p'' and
``n'' and the shear modulus should obviously be set to zero.  

To complete the dynamical equations, we need to provide 
the quantity $\varepsilon_{\x}$, that accounts
for the entrainment, and the ``external'' force density $f^\x_i$  that acts on
each fluid component.  As we will see later, the force can be used to account for the 
crust elasticity (and, in general, also the magnetic field). 
Finally, the system of equations is closed by an equation of state
(EoS). It can be described by an energy functional that ensures Galilean invariance;
\begin{equation}
\mathcal{E} = \mathcal{E} \left( \rho_\x,  w_{\x \y}^2
\right) \, . \label{eq:EoS}
\end{equation}
The chemical potential $\tilde{\mu}_\x$ and the entrainment parameter
$\varepsilon_{\x}$ are then determined from \footnote{ In~\cite{2009MNRAS.396..951P},  the definition of the entrainment parameter $\veps_{\x}$ given 
in equation~(7) contains a typo. Here, we provide in  equation~(\ref{eq:vareps}) the correct definition. } 
\begin{eqnarray}
\tilde{\mu}_{\x} & \equiv & \left. \frac{\partial \mathcal{E}}{\partial \rho_{\x} }
\right| _{\rho_{\y}, w_{\x\y}^2}\, , \label{eq:defmu} \\
\rho _{\x}  \varepsilon_{\x} & \equiv & 2 \left.  \frac{\partial
\mathcal{E}}{\partial w^{2}_{\x\y} } \right|_{\rho_\x,\rho_\y} \, . \label{eq:vareps}
\end{eqnarray}

\subsection{Equilibrium model}

Let us first consider the equations that need to be solved in order to determine a 
non-rotating superfluid neutron star in dynamical and chemical equilibrium.
We consider a static, nonrotating, background star in which matter is in
$\beta$-equilibrium, which means that $\tilde \mu \equiv \tilde \mu_{\p} = \tilde
\mu_{\n}$. Moreover, we assume that the crust is unstrained, which means that the elasticity 
affects only the linear perturbations of the system. 
Under these assumptions, the star is described by the
following equations:
\begin{eqnarray}
\nabla \tilde \mu & = & - \nabla \Phi \, ,  \label{eq:bg1} \\
\nabla^2 \Phi & = & 4 \pi G \rho \,  . \label{eq:bg2} 
\end{eqnarray}
Equation~(\ref{eq:bg1}) is
obtained from the stationary Euler equations, the $\beta$-equilibrium
condition and the standard relation between the pressure and chemical
potential $\nabla P = \rho \nabla \tilde \mu $. The system of
equations~(\ref{eq:bg1}) and~(\ref{eq:bg2}) can be written as a single
ordinary differential equation
\begin{equation}
\nabla^2 \tilde \mu =  - 4\pi G \rho \,  .   \label{eq:bg}
\end{equation}
The background configuration is therefore completely determined once we
choose an EoS that provides a second relation between the mass density
and the chemical potential. More details on the EoSs used in this work
are given below, in Sec.~\ref{sec:EoS}. 

\subsection{Lagrangian perturbations}

In order to model the dynamics of the crust, where the elastic restoring force
depends on the deviation from the unstrained configuration, it is natural to turn to Lagrangian perturbation theory. 
In the problem considered in this paper, the relevant perturbations are trivially related to 
the Eulerian variations of the various quantities, but in the general (rotating and possibly pre-strained) case
the Lagrangian description becomes more involved. However, there has been recent progress in developing the required 
theory~\citep{2011arXiv1105.1244A}, and we draw on these results in the following. In fact, our analysis provides the first application of the new formalism. 

In the two-fluid situation, it is natural to introduce a Lagrangian displacement
vector $\xi_{\x}^i$ for each fluid
component~\citep*{2004MNRAS.355..918A}.  In order to describe
the motion associated with these vectors, we define the
Lagrangian perturbation of a general (scalar or vector) quantity $Q$ along the vector
field $\xi_\x^i$,
\begin{equation}
\Delta_\x Q = \delta Q + \mathcal{L}_{\xi_{\x}} Q \,  , 
\end{equation}
where $\delta Q$ is the Eulerian perturbation of $Q$ and
$\mathcal{L}_{\xi_{\x}} $ denotes the Lie derivative along $\xi_\x^i$.

The dynamics of neutron star oscillations can be studied by
linearising the system of
equations~(\ref{eq:mconv})-(\ref{eq:Pois}). The linearised mass
conservation and Poisson equations are then given by;
\begin{eqnarray}
&&  \Delta_{\x} \rho_\x + \rho_\x \nabla_{i} \xi^{i}_{\x} = 0 \, ,  \label{eq:Cm}\\
&&  \nabla^2 \delta \Phi = 4 \pi G \left( \delta \rho_{\x} + \delta
  \rho_{\y} \right)  \, . \label{eq:P}
\end{eqnarray}
Meanwhile, for non-rotating stars, the linearised momentum conservation equations
take the form~\citep{2009MNRAS.396..894A}
\begin{eqnarray}
&&  \left( 1 - \veps_\ch \right) \frac{ \partial^2 \xi _{i}^{\ch} }{\partial t ^2 } + \veps_\ch \frac{ \partial^2 \xi _{i}^{\f}
}{\partial t ^2 } + \nabla_{i} \delta \Phi
                    + \xi_{\ch}^{j} \nabla_{j}\nabla_{i} \Phi - \left( \nabla_{i} \xi_{\ch}^{j} \right) \nabla_{j} \tilde \mu_{\ch} 
                     + \nabla_{i} \Delta_{\ch} \tilde \mu_{\ch}  =  \frac{1}{\rho_\ch} \nabla^{j} \sigma_{ij} \, ,  \label{eq:eul-1} \\
&&  \left( 1 - \veps_\f \right)  \frac{ \partial^2 \xi _{i}^{\f}
}{\partial t ^2 }  + \veps_\f \frac{ \partial^2 \xi _{i}^{\ch}
}{\partial t ^2 } + \nabla_{i} \delta \Phi
                     + \xi_{\f}^{j} \nabla_{j}\nabla_{i} \Phi - \left( \nabla_{i} \xi_{\f}^{j} \right) \nabla_{j} \tilde \mu_{\f} 
                     + \nabla_{i} \Delta_{\f} \tilde \mu_{\f} = 0 \, ,  \label{eq:eul-2} 
\end{eqnarray}
where the elastic stress tensor is defined as
\begin{equation}
  \sigma_{ij} = \check \mu \left( \nabla_{i} \xi_{j}^{\ch} + \nabla_{j}
    \xi_{i}^{\ch} \right) - \frac{2}{3} \check \mu \left(
    \nabla^{k}\xi_{k}^{\ch} \right) \delta_{ij} \,  .
\end{equation}
The shear modulus is denoted by $\check \mu$, and must not be confused with
the chemical potential~(\ref{eq:defmu}). 
For nonrotating configurations in $\beta$-equilibrium,
equations~(\ref{eq:eul-1}) and~(\ref{eq:eul-2}) assume the
simpler form;
\begin{eqnarray}
&&  \left( 1 - \veps_\ch \right)  \frac{ \partial^2 \xi _{i}^{\ch} }{\partial t ^2 }  + \veps_\ch \frac{ \partial^2 \xi _{i}^{\f}
}{\partial t ^2 }+ \nabla_{i} \left( \delta \tilde \mu_{\ch} + \delta \Phi  \right) 
                                                       =  \frac{1}{\rho_\ch} \nabla^{j} \sigma_{ij} \, ,  \label{eq:eul-3} \\
&&  \left( 1 - \veps_\f \right) \frac{ \partial^2 \xi _{i}^{\f}
}{\partial t ^2 } + \veps_\f  \frac{ \partial^2 \xi _{i}^{\ch} }{\partial t ^2 } + \nabla_{i} \left( \delta \tilde \mu_{\f}  + \delta \Phi \right) 
                     = 0 \, .  \label{eq:eul-4} 
\end{eqnarray}
As noted earlier, Equations~(\ref{eq:eul-3}) and~(\ref{eq:eul-4}) also describe the
dynamics of the neutron star core provided we replace, respectively,  the indices c
and f with p and n, and set $\check\mu$ to zero.

\subsection{Chemical gauge and crust/core junction conditions}  \label{sec:CCbc}

In the inner neutron star crust, the protons and a sizeable fraction of the neutrons are bound by the
strong interaction. They form a lattice of nuclei that become heavier
towards the crust/core transition, and which supports elastic stresses.
The remaining fraction of (unbound) neutrons form a superfluid that may
flow through the confined nucleons. This superfluid component
first appears at the neutron drip density, $\rho_{\rm{ND}} =
4.3\times10^{11}~\textrm{g cm}^{-3}$, which marks the transition
between the inner and outer crust, where there are no free neutrons.

In the inner crust there are, at least, two dynamical degrees
of freedom; a confined component of protons and neutrons and the free
superfluid neutrons. However, the distinction
between superfluid and confined neutrons is not obvious. It depends on
the local density and the dynamical time-scale of the
process under consideration. Adopting the operational definition 
due to~\cite{2006MNRAS.368..796C}, a neutron may be considered
free when (on the time-scale considered) its energy is sufficient to
overcome the potential barriers separating the nuclei,
either classically or by quantum tunnelling. At the bottom of the
crust, where the nuclei become denser and form exotic structure
(the so-called pasta phase), the potential wells become closer and marginally bound
states may exist.  These states may penetrate the potential barriers on
time-scales that are macroscopically long (but cosmologically short),
see ~\cite{2006MNRAS.368..796C} for discussion.

Within the two-fluid formalism, this uncertainty may be expressed in terms of the
chemical gauge parameter $a_c$ introduced by~\citep{2006CQGra..23.5367C,2006MNRAS.368..796C,2011arXiv1105.1244A}:
\begin{equation}
n_{\f} = n_{\n} + \left( 1 - a_\ch \right) n_{\p} \, , 
\qquad \qquad n_{\ch} = a_{\ch} n_{\p} \, ,
\end{equation}
where $n_{\f}$ and $n_{\ch}$ are, respectively, the number density of
the free neutrons and the confined component, while $n_{\n}$ and
$n_{\p}$ are the \underline{total} number densities of neutrons and protons.
One can show that the neutron conjugate momentum and the
chemical potential are independent of the chemical
gauge choice~\citep{2006CQGra..23.5367C,2006MNRAS.368..796C,2011arXiv1105.1244A} as long
as $a_{\ch}$ is either held fixed or depends only on the nuclear charge
number $Z$.  This gauge independence is important for the
derivation of the crust/core junction conditions.

In order to study the oscillations of a given neutron star model, we must prescribe
boundary conditions at the star's origin and surface. In a model with a
crust, we must also specify conditions at the crust/core interface
and at the outer/inner crust transition.  In this work, we implement  the
static limit of the crust/core conditions derived by~\cite{2011arXiv1105.1244A}.
From the mass conservation equations, we obtain the conditions for
the radial component of the Lagrangian displacements at $r=R_{cc}$:
\begin{eqnarray}
\xi_{\ch}^{r} & = & \xi_{\p}^{r} \, ,  \label{eq:bc1} \\
\xi_{\f}^{r}  & = & \frac{x_{\p}-x_{\ch}}{1-x_{\ch}} \xi_{\p}^{r} +
\frac{1-x_{p}}{1-x_{\ch}} \xi_{\n}^{r} \,  . 
\end{eqnarray}
As in the single fluid case the other junction conditions may
be derived from the total momentum conservation equation, which
leads to the following vertical and horizontal
projections~\citep{2011arXiv1105.1244A}:
\begin{eqnarray}
&& \langle \Delta_{\ch} P + \frac{2}{3} \check\mu \nabla_{i} \xi^{i}_{\ch} -
2 \check \mu \, \frac{d \xi _{\ch}^{r}}{d r} \rangle = 0  \, ,   \label{eq:tr1} \\
&&  \left( g^{i j} -  N^{i}  N^{j}  \right)  \langle \check \mu 
N^{k} \left(  \nabla_{j} \xi_{k}^{\ch} + \nabla_{k} \xi_{j}^{\ch}
\right)  \rangle  = 0 \,  .  \label{eq:tr2}
\end{eqnarray}
where $g_{i j}$ is the (flat) metric, $ N ^{i}$ is a unit vector
orthogonal to the crust/core interface and $\langle \dots \rangle$
denotes the change of a physical quantity across the
interface.  Equations~(\ref{eq:tr1}) and~(\ref{eq:tr2}) reduce to the
standard traction conditions in the single fluid case. Finally, from
the chemical gauge independence we can determine the condition for the
chemical potential of the superfluid neutron components;
\begin{equation}
\Delta_{\ch} \tilde \mu _{\f} = \Delta_{\p} \tilde \mu_{\n} \, . \label{eq:bcmu}
\end{equation}
From the junction condition~(\ref{eq:bc1}), it is clear that for
nonrotating models in $\beta$-equilibrium equation~(\ref{eq:bcmu}) is
equivalent to $\delta \tilde \mu_{\f} = \delta \tilde \mu_{\n}$.

These conditions complete the description of the problem that needs to be solved in order to 
determine that star's oscillation spectrum (see Sec.~\ref{sec:polbc} for the inner/outer crust boundary conditions).

\section{Model Equation of State}  \label{sec:EoS}

In order to make maximal use of the theoretical framework, we ought to build our models using a realistic equation of state for supranuclear matter. 
We will, however, resort to using a simple model EoS.  There are several reasons for this. 
Most importantly, we do not yet have a realistic EoS that provides all the different parameters required in our analysis. 
Most tabulated EoS simply provide the pressure vs density relation, whereas we need (at the very least) 
detailed information about the composition, the 
superfluid pairing gaps and the entrainment between neutrons and protons. Progress in this direction is being made, c.f.,~\citet{2008MNRAS.388..737C}, but the models are not yet complete. Still, the developments have reached the point where it would be meaningful to consider a fully relativistic analysis (as is needed to make the use of a realistic EoS meaningful).
This is, however, beyond the scope of the present work. 
Our main interest here is to probe the key issues and consider the associated phenomenology. 
For this exercise it makes sense to focus on
a simple analytic model with freely adjustable parameters. 

We model the neutron star matter in terms of two
polytropic EoSs. Despite the simplicity of the construction, combinations of
polytropes
enable us to study the effects of composition stratification and
symmetry energy on the oscillation spectrum. Moreover, we have already used
these EoSs in previous work~\citep{2011MNRAS.413...47P,2010MNRAS.405.1061S}, which means that we can build on, and compare to, the 
  previous results. As a new application of
the polytropic EoS, we construct  stratified two-fluid models with
different parameters in the crust and the core in order to
approximate the qualitatively different compositions in these two
regions. 

In most astrophysical systems the relative velocity between the two
fluids is small. Therefore, equation~(\ref{eq:EoS}) can be expanded in
a series:
\begin{equation}
\mathcal{E} = \mathcal{E}_{0} \left(\rho_\f, \rho_\ch \right)
+ \alpha_0 \left( \rho_\f, \rho_\ch \right) w_{\f \ch}^2 + \mathcal{O}\left(w_{\f \ch}^4\right) \, , \label{eq:EoSbulk}
\end{equation}
This approximation has the 
advantage that the bulk EoS, $\mathcal{E}_{0}$, and the
entrainment parameter, $\alpha_0$, can be independently specified at
 $w_{\f\ch}^i=0$.  From equation~(\ref{eq:vareps}) it follows
that the entrainment parameter $\varepsilon _{\x}$ is related to the
function $\alpha_0$ by
\begin{equation}
\rho_\x  \varepsilon_\x = 2 \alpha_0 \, .  \label{eq:alp}
\end{equation}
In a co-moving background, the density perturbations can be expressed in
terms of the chemical potential perturbations as:
\begin{eqnarray}
\Delta \rho_\f  & = & \mathcal{S}_{\f\f}  \Delta \tilde \mu _\f + \mathcal{S}_{\f\ch}  \Delta \tilde \mu _\ch \, , \\
\Delta \rho_\ch & = & \mathcal{S}_{\ch\f} \Delta \tilde \mu _\f + \mathcal{S}_{\ch\ch} \Delta \tilde \mu _\ch \, ,
\end{eqnarray}
where 
\begin{equation}
\mathcal{S}_{\x\y} \equiv \frac{\partial \rho_\x}{\partial \tilde \mu_\y} \, . \label{eq:Sdef}
\end{equation}

\subsection{The bulk equation of state}

Given the above relations, we need to first of all provide the energy $\mathcal{E}_0$.
The first EoS we consider has also been used by~\citet{2002A&A...381..178P,2004MNRAS.347..575Y,
  2009MNRAS.396..951P}:
\begin{equation}
\mathcal{E}_{0} =   \frac{K}{1-\left( 1 + \sigma_{sym} \right) x_\ch} \rho_\f ^2
- \frac{2 K \sigma_{sym} }{1-\left( 1 + \sigma_{sym} \right) x_\ch} \rho_\f \rho_\ch
+ \frac{ K  \left[ 1 + \sigma_{sym} - \left( 1 + 2 \sigma_{sym} \right) x_\ch \right] }
{x_\ch \left[  1-\left( 1 + \sigma_{sym} \right) x_\ch \right] } \rho_\ch^2 \label{eq:EoSA} \, ,
\end{equation}
where $K$ is a polytropic constant, $x_{\ch}$ is the confined nucleon fraction
and $\sigma_{sym}$ is a parameter that can be related to the symmetry
energy~\citep{2002A&A...381..178P}. In this EoS, both $x_{\ch}$ and
$\sigma_{sym}$ are taken to be constant~\citep{2009MNRAS.396..951P}.  
Using equation~(\ref{eq:Sdef}) and (\ref{eq:EoSA}), we readily obtain
\begin{eqnarray}
\mathcal{S}_{\ch\ch} & = & \frac{ x_\ch }{ 2 K \left( 1 + \sigma_{sym} \right) } \, , \\
\mathcal{S}_{\f\f}   & = & \frac{ 1+\sigma_{sym} - \left( 1 + 2
                             \sigma_{sym} \right) x_\ch }{ 2 K \left( 1 + \sigma_{sym} \right) } \, , \\
\mathcal{S}_{\ch\f}  & = & \mathcal{S}_{\f\ch} = \sigma_{sym} \, \mathcal{S}_{\ch\ch} \, .
\end{eqnarray}
As discussed by~\cite{2009MNRAS.396..951P}, this EoS leads to non-stratified stellar models.
The global properties of these models are similar to those of
the $N=1$ polytrope in the single fluid case. For instance,
the pressure is related to the total mass density by the standard
equation $P=K \rho^2$. 
The set of models constructed from this EoS will be referred to as models A,  c.f. 
\cite{2009MNRAS.396..951P}.

At this point, it is useful to pause and make some comments on the numerical 
implementation and the resultant stellar models.
In our numerical code,
we use dimensionless units based on the gravitational constant $G$,
the central mass density $\rho_{0}$ and the stellar radius $R$.  Unless 
directly specified, all models A  considered in this paper are such that the crust/core
transition is at $R_{cc} = 0.9 R $. Therefore, from the expected crust/core
transition density $\rho_{cc} = 1.2845\times10^{14}~\text{g
  cm}^{-3}$~\citep{2001A&A...380..151D}, we can determine the central
mass density of the star. At $r=R_{cc} $ the dimensionless mass
density is $\rho_{cc} = 0.109~\rho_{0}$, hence $\rho_{0}=1.175\times
10^{15}~\text{g cm}^{-3}$. Meanwhile, the dimensionless mass of models A is
$M/(\rho_{0} R^{3} ) = 1.273$. Therefore, the physical mass and
the radius can also be determined. For instance, in the case of the canonical neutron star
mass, $M=1.4~M_{\odot}$, we obtain the radius
$R=12.29~\textrm{km}$.

It is obviously not the case that the component fractions are
constant in a real neutron star. We can account for composition gradients
by considering another simple combination of polytropes;
\begin{equation}
\mathcal{E}_{0} = k_\f \rho_\f ^{\gamma_\f} + k_\ch \rho_\ch
^{\gamma_\ch} \, , \label{eq:EosPR}
\end{equation}
where $k_\x$ and $\gamma_\x$ are constants. This EoS does not have
the symmetry energy term, but it can leads to models with varying composition
when $\gamma_\f \ne \gamma_\ch$.
From equations~(\ref{eq:defmu})
and~(\ref{eq:EosPR}) it follows that the chemical potential and the corresponding mass
density  are related by
\begin{equation}
\rho_\x = \left( \frac{\tilde \mu _\x }{k_\x \gamma_\x } \right)
^{N_\x} \, ,   \label{eq:rhomu}
\end{equation}
where the polytropic index is given by $N_\x = \left( \gamma_{\x} -1
\right) ^{-1}$. From this result we can determine the proton fraction for a given stellar
 model by imposing  $\beta$-equilibrium. After some calculations, we obtain:
\begin{equation}
x_\ch = \left[ 1 + \frac{ \left( \gamma_\ch k_\ch \right) ^{N_\ch}
}{\left( \gamma_\f k_\f \right) ^{N_\f}} \, \tilde \mu ^{N_\f-N_\ch}
\right]^{-1} \, .  \label{eq:xp}
\end{equation}
For $\tilde \mu \to 0$, it is clear from equation~(\ref{eq:xp}) that
$x_\ch$ vanishes for $N_\f > N_\ch$ and tends to unity when $N_\f <
N_\ch$. 
\begin{figure}
\begin{center}
\includegraphics[height=75mm]{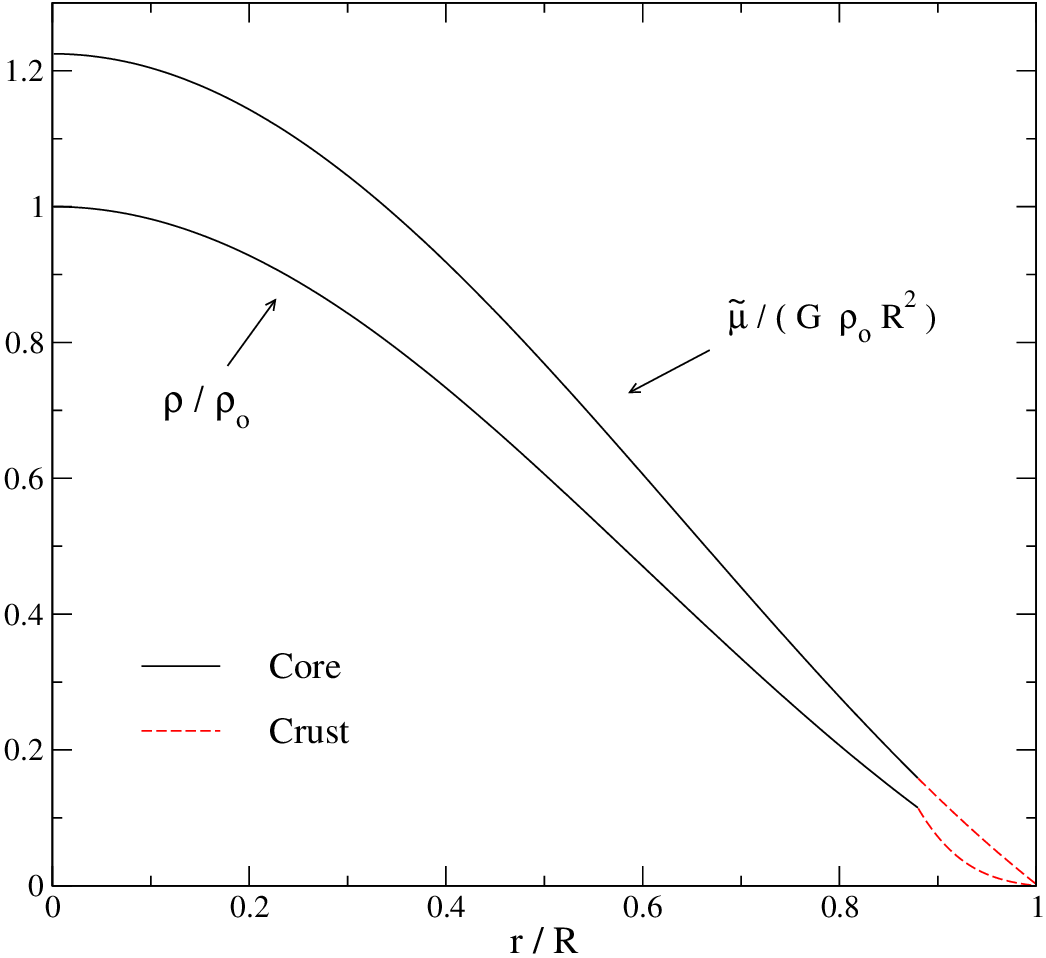}
\includegraphics[height=75mm]{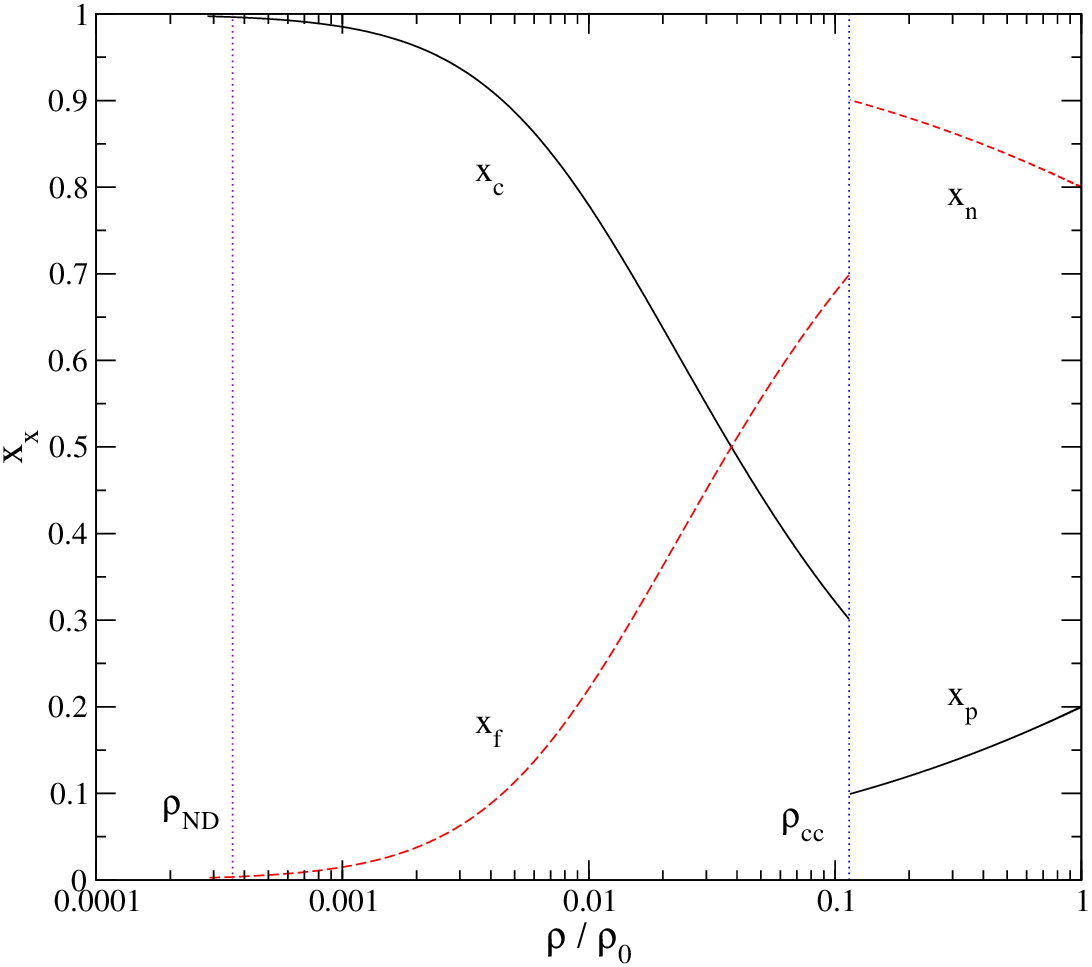}
\caption{This figure shows some properties of a two fluid polytropic
  model D with $x_{\ch} = 0.3$ at $r=R_{cc}$. The left panel displays
  the radial profile of the mass density and the chemical
  potential. The right panel shows (on a semi-logarithmic scale) the
  fraction of the crust/core constituents (vertical axis) and the
  total mass density (horizontal axis). In the right panel the
  vertical dotted lines denotes the density of the crust/core
  transition $\rho_{cc}$ and the neutron drip density
  $\rho_{\rm{ND}}$.  Note that the total number of neutrons and protons is
  continuous at  $r=R_{cc}$, as in the crust a fraction of  neutrons is
  confined in the nuclei of the lattice.  
  \label{fig:bm1}}
\end{center}
\end{figure}
\begin{table}
\begin{center}
  \caption{\label{tab:D} This table provides the main parameters for
    the stellar model sequence D. The core is described by the
    EoS~(\ref{eq:EosPR}) with $N_\n = 1$, $N_\p = 1.4$ and $ k_\n =
    0.766$, $k_\p = 2.256$. The crust/core transition has been set to
    $R_{cc} = 0.88 R$. In the crust, we choose the same polytropic
    indices for the free neutrons and confined nucleons, respectively,
    $N_\f = 2.9$ and $N_\ch =1.2$, but vary $x_{\ch}$ at
    $r=R_{cc}$. For different values of $x_{\ch}$ at the crust/core
    interface (first column), we show in the second and third column
    the dimensionless EoS parameters $k_\ch$ and $k_\f$,
    respectively. The dimensionless mass is given in the fourth
    column. Note that the coefficients $k_\x$ are given in units of $G
    R^{2} \rho_{0}^{2-\gamma_\x}$}
\begin{tabular}{  c c c c   }
  \hline
   $ x_{\ch} $  &  $ k_{\ch} $ & $ k_{\f}$  & $M /\left( \rho_{o} R^3 \right)$ \\
   \hline
    0.1   &  3.574  & \, 0.258 & 1.149 \\
    0.3   &  1.431  & \, 0.281 & 1.155 \\
    0.5   &  0.935  &  \, 0.316 & 1.161 \\
    0.7   &  0.706  &  \, 0.377 & 1.167 \\
    1.0   &  0.525  & 13.172 & 1.176 \\
\hline
\end{tabular}
\end{center}
\end{table}

In a ``realistic'' neutron star, the proton fraction decreases from
the centre to the crust/core interface, while the confined component
increases from the bottom of the crust and reaches unity at the
neutron drip. This behaviour can be approximated using the
EoS~(\ref{eq:EosPR}) provided we choose different values for the
polytropic indices in the core and the crust, i.e. let $(N_\n,N_\p)\neq
(N_\f,N_\ch)$. For simplicity, we also assume that the equilibrium star is such that 
the total
mass density, the chemical potential and its first derivative are
continuous at the crust/core interface.
We have found a good compromise that satisfies all these conditions by
choosing  $N_\n = 1$, $N_\p = 1.4$ and $ k_\n = 0.766$, $
k_\p = 2.256$, in the core and   $N_\f = 2.9$, $N_\ch =
1.2$ and $ k_\f = 0.281$, $ k_\ch = 1.431$ in the crust. The crust/core transition
has been fixed at $R_{cc} = 0.88 R$.  Note that the coefficients
$k_\x$ are given in units of $G R^{2} \rho_{0}^{2-\gamma_\x}$.  With these
parameters, the proton fraction in the core ranges from $x_\p = 0.2$
at $r=0$ to $x_\p = 0.1$ at the crust/core boundary. These values are on the high side, but this is not too concerning here. In the crust, the
charged component varies from $x_\ch = 0.3$ at $r=R_{cc}$ to $x_\ch =
1$ at the inner/outer crust interface, $r=R_{\rm{ND}}$. In this model,
the total mass density at the interface is  $\rho_{cc} =
0.115~\rho_{0}$ in dimensionless units. As in the case of model A, we can therefore
determine the central mass density, $\rho_{0} =
1.117\times10^{15}~\text{g cm}^{-3}$. For a star with mass
$M=1.4~M_{\odot}$, we can determine the stellar radius from the
dimensionless mass of this model; from $M/(\rho_{0} R^{3} ) = 1.155$ we find that the radius is
$R=12.92~\textrm{km}$.

Fig.~\ref{fig:bm1}  illustrates the properties of this crust-core model. The mass
density and the chemical potential are shown in the left panel and the
constituent fractions are provided in the right panel. The crust/core and  
neutron drip transitions are also indicated.  We have
assumed a neutron drip density $\rho_{\rm{ND}} = 4.3\times
10^{11}~\text{g
  cm}^{-3}$~\citep{1983bhwd.book.....S,2001A&A...380..151D} which for
this stellar model corresponds to $R_{\rm{ND}}=0.9988 R$.
For the same core model, we have also studied solutions with different values of
 $x_{\ch}$ at the crust/core interface.  From 
equation~(\ref{eq:xp}) and the continuity conditions for the total
mass density and chemical potential, it is evident that the
coefficients $k_\ch$ and $k_\f$ depend on the parameter $x_{\ch}$ at
$r=R_{cc}$.  Their values and the total mass of the stellar model are
listed in Table~\ref{tab:D}.  We refer to this sequence of stratified
stars as models D in order to distinguish them from models B and C
considered in previous
work~\citep{2009MNRAS.396..951P,2010MNRAS.405.1061S,2011MNRAS.413...47P}.

\begin{figure}
\begin{center}
\includegraphics[height=75mm]{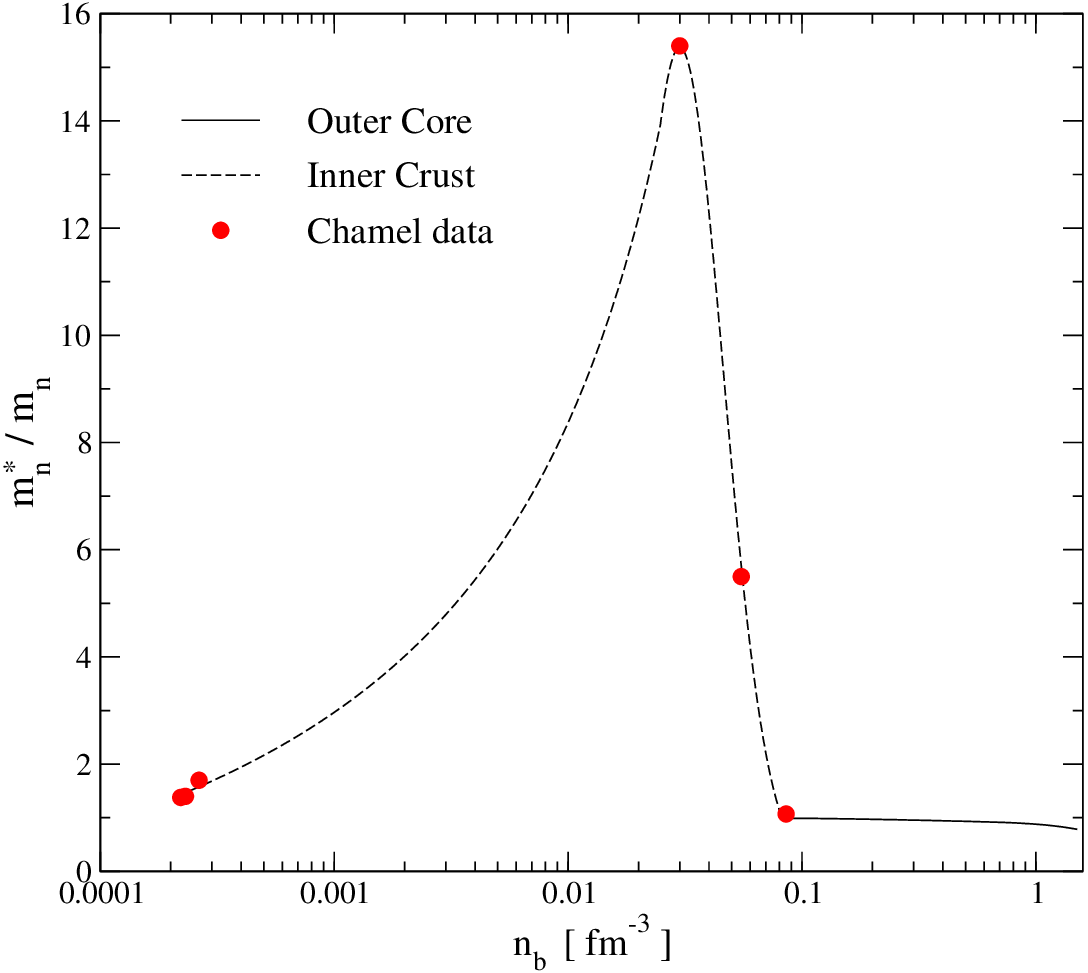}
\includegraphics[height=75mm]{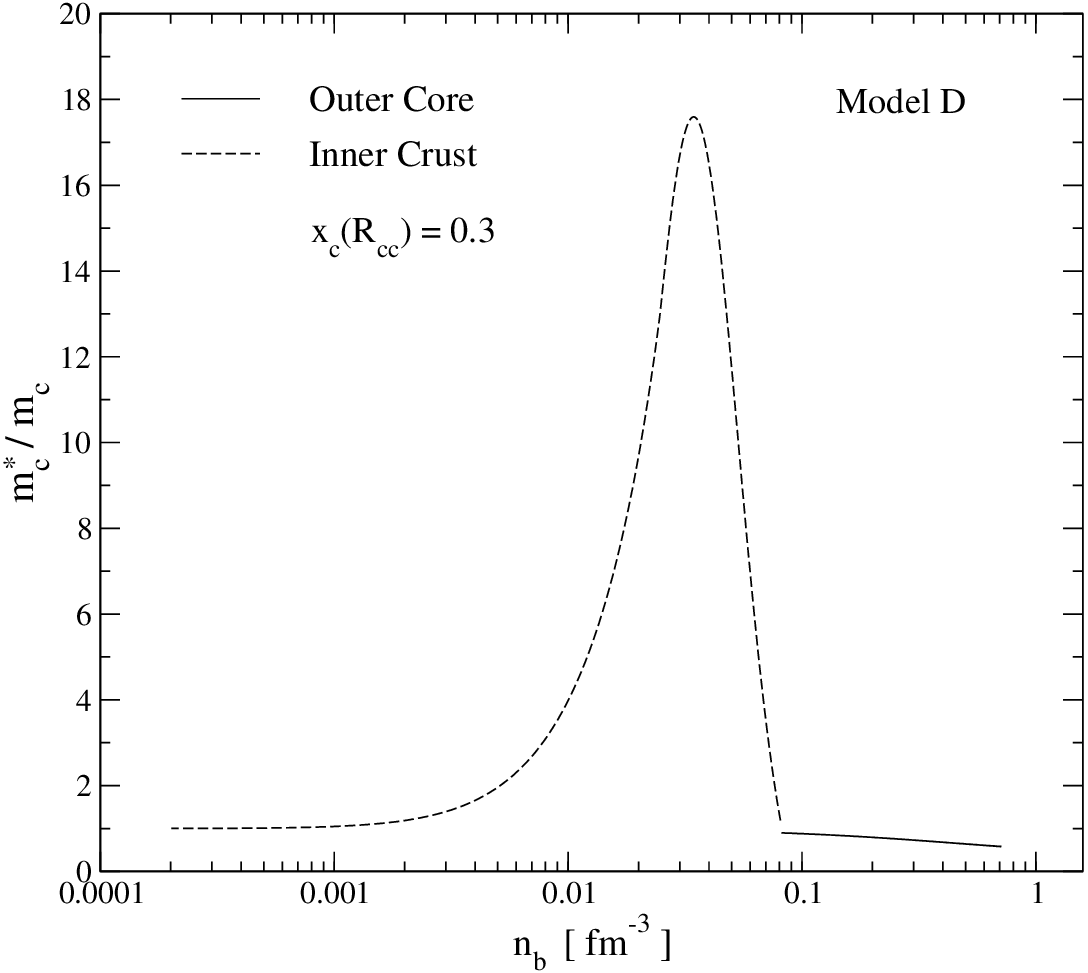}
\caption{ This figure shows the dependence of the effective masses on
  the baryon number density. In the left panel, we show the neutron
  effective mass for both the core (neutrons) and the crust (free
  neutrons). In the crust, the filled circles denote the data set
  determined by Chamel and
  collaborators~\citep{2005NuPhA.748..675C,2005NuPhA.747..109C,2006NuPhA.773..263C}.
  These data have been fitted with the curve (dashed-line) shown in
  the left panel. For model D with $x_{\ch} (R_{cc} ) = 0.3$, we
  display in the right panel the effective mass for the charged
  component (core) and the confined particles
  (crust).  \label{fig:bm2} }
\end{center}
\end{figure}

\subsection{Entrainment}   \label{sec:entr}
%
In a real neutron star, superfluid neutrons may flow through the lattice of
nuclei in the crust and also coexist with superconducting protons in
the core. The different constituents are coupled through
entrainment, mutual friction, crust/core boundary conditions, EoS,
etc.
The entrainment is a non-dissipative process that arises from strong interaction. Owing to this effect each
superfluid constituent can carry along a fraction of the
other component. From a dynamical point of view, it is natural to account
for the entrainment by introducing an effective mass associated with  
each component of the system and which is in general
different from the bare particle mass. Meanwhile, in the two-fluid formalism, the
entraiment is usually described by the parameter $\veps_\x$ 
defined by equation~(\ref{eq:vareps}). It is related to the effective
mass $m_\x^\star $ as
\begin{equation} \veps_\x = 1 - \frac{m_\x
  ^\star}{m} \, .  \label{eq:enteff}
\end{equation}
Entrainment affects the constituent conjugate momentum in such a way that it is not longer aligned
with the transport velocity of the fluid constituent.

In recent work, \cite{2008MNRAS.388..737C} determined the nucleon
effective masses in a neutron star core. The results are based on the
two-fluid formalism and describe the strong interaction between
nucleons using a two-body force of the Skyrme type with different
parametrisations. In a non-relativistic
two-fluid model the effective masses can  be described by the following
relation~\citep{2008MNRAS.388..737C}:
\begin{equation} 
m^{\star}_{\x} =  m_{\x}  \frac{ 1 + \hat \beta_{3} \rho_\x}{ 1+  \hat
  \beta_{3}  \rho_\ba}   \, ,  \label{eq:meff_n}
\end{equation}
where $\rho_{\ba}$ is the baryon mass density, and 
$\hat \beta_3 $ is defined by
\begin{equation}
\hat \beta_{3} = \frac{\beta_{3} }{ m} = \frac{2 B_3 }{\hbar^2}  \,  ,
\end{equation}
where $\hbar$ is the reduced Planck's constant.  The quantity $B_3$ is
a coefficient that depends on the parametrisation of the Skyrme
force~\citep{2008MNRAS.388..737C}. 

In this work, the superfluid neutron star core is modeled by two fluid
polytropes and the effective neutron mass is determined
by equation~(\ref{eq:meff_n}). For the proton effective mass, we 
use the EoS constraint~(\ref{eq:alp}), i.e. 
\begin{equation}
\rho_\n \veps_\n = \rho_\p \veps_\p  \, , \label{eq:EoScon}
\end{equation}
together with the definition~(\ref{eq:enteff}). This model is obviously not
consistent, as the effective mass relation~(\ref{eq:meff_n}) has been
determined for ``realistic'' EoS and not for polytropes.
However, this is the first time that the complete  oscillation spectrum of
superfluid neutron star with an elastic crust is considered. Hence, it is natural 
to focus on the phenomenology sacrificing (to some extent)  realism.
Our stellar model has the key features expected of a real neutron star, but the ``proportions'' may not be quite 
realistic. This should, however, be  easy to fix once the problem is considered in General Relativity.
The relevant developments are, in fact, already under way.

The effective mass of the free superfluid neutrons in the inner crust has
been determined within the band theory approximation, but only for a few
specific
densities~\citep*{2005NuPhA.748..675C,2005NuPhA.747..109C,
  2006NuPhA.773..263C}. The results indicate that, in the deep crust, the effective mass can be
as large as $m^{\star}_{\f} = 15.4\, m_{\f}$. 
As already discussed by~\cite{2009MNRAS.396..894A}, who studied the
problem within a local (plane-wave) approximation, these large entrainment
values may have considerable effect on the properties of the oscillation
spectrum.  The currently available entrainment data is shown in the left panel
of Fig.~\ref{fig:bm2}. We interpolate these data points with a curve,
and determine the proton effective mass with
equation~(\ref{eq:EoScon}) by applying the method that was used for the
core.  The results are shown in Fig.~\ref{fig:bm2}, where the right
panel displays the proton effective mass for a model D with $x_\ch =
0.3$ at the crust/core interface.


\section{The eigenvalue problem} \label{sec:EP}

For non-rotating stellar models, the study of the linearised
equations~(\ref{eq:Cm})-(\ref{eq:P}) and
(\ref{eq:eul-3})-(\ref{eq:eul-4}) is considerably simplified if we
expand the perturbation variables in vector harmonics. For each set of
harmonic indices $(l,m)$ (associated with the familiar spherical harmonics $Y_l^m$), the relevant equations then depend only on the radial
coordinate $r$ and time $t$. In order to set up an eigenvalue problem, we
assume that the perturbation variables have a harmonic time
dependence $e^{ \mathrm{i} \sigma t}$ (this essentially amounts to taking a Fourier transform of the equations). The mode eigenfrequency $\sigma$ is in
general a complex number, where the real part describes the oscillation frequency
and the imaginary part the growth/damping time of the
mode under consideration. However, in this work the eigenfrequency $\sigma$ is real since we 
are not accounting for  dissipative processes
or mechanisms that may  trigger instabilities.

The perturbations can be decomposed into two
classes of different parity, namely the polar (spheroidal) and the
axial (torsional) perturbations. The spectrum of non-rotating stars is also
degenerate with respect to the azimuthal harmonic $m$.  This degeneracy
is removed when we account for additional physics like 
rotation or the star's magnetic field.  Oscillation modes of a
non-rotating star are commonly classified by the dominant restoring
force that acts on the displaced fluid elements, the harmonic index
$l$ and the radial number $n$~\citep{Cowling:1941co} that labels the ``overtones'' of a given set of oscillations.
In simple models the latter parameter can be associated with the number of radial nodes in the 
relevant eigenfunctions, but this identification no longer holds for more complex systems.

The axial part of the oscillation spectrum is comparatively simple.  For
non-rotating stars without magnetic fields and crust, the axial
perturbations form a degenerate set of zero frequency modes.  The crust
elasticity breaks this degeneracy as it sustains a set of torsional
shear modes.  We denote the torsional modes (t-modes) by the
symbol ${}^l \textrm{t}_{n}$, where $l$ is the harmonic index and $n$ labels the overtones.
 For any $l$, the frequency of the torsional modes
assumes its lowest value for the fundamental torsional
 mode and increases with $n$. In compact stars, the
typical frequency of the ${}^{2} \textrm{t}_{0}$-mode is a few tens of Hz~\citep{1988ApJ...325..725M}.

The polar oscillation spectrum is more complex. In
addition to the shear waves, the presence of pressure and
composition/thermal gradients may generate acoustic and gravity waves. The acoustic
modes are mainly
restored by pressure variations (p-modes) and cover the high frequency
band of the spectrum, above 1~kHz. At lower frequencies, typically
below 100~Hz, composition and thermal gradients can sustain the class
of gravity modes (g-modes), which are restored by buoyancy.  For a
given $l$, there is an infinite series of pressure (gravity) modes
with $n \ge 1$, whose frequency is increasing (decreasing)
with $n$.  The $\xi^{r}$ eigenfunction of the pressure, ${}^l
\textrm{p}_{n}$, and gravity modes, ${}^l \textrm{g}_{n}$, have $n$ radial
nodes.  The fundamental mode (f-mode), ${}^l \textrm{f}$, separates the
class of p- and g-modes. Its frequency scales with the averaged
stellar density and no radial nodes are present in its $\xi^{r}$
eigenfunction.

The crust elasticity affects the polar oscillation  
spectrum in two ways. Firstly, it generates a class of spheroidal shear modes
(s-modes) and secondly it slightly modifies the acoustic mode
frequencies~\citep{1988ApJ...325..725M}.  Spheroidal shear modes are
represented by ${}^l \textrm{s}_{n}$, with $n\ge 1$. Their frequencies  
increase with $n$, and for small $n$ they lie below the
f-mode frequency.  In the very low frequency band of the spectrum,
there are also interface modes,which are denoted by ${}^l
\rm{i}$. These modes arise from the presence of internal interfaces and
they
strongly depend on the physical conditions near these interfaces. For
any $l$, there is an ${}^l \rm{i}$ mode associated with each internal interface.

The mode classification presented above allows us to describe the complete 
oscillation spectrum of  non-rotating, non-magnetised
single-fluid stars.  In superfluid neutron stars, the multi-component dynamics of
the system lead to additional degrees of freedom and the
oscillation spectrum is therefore richer.
In fact, in a two-fluid model we can (usually) decompose the
dynamics into a co-moving (counter-moving) motion, where the fluid
elements oscillate in (and out of) phase. As a result, we can distinguish
two classes of acoustic modes that are commonly called ``ordinary''
and ``superfluid'' modes. The ordinary modes have spectral properties
similar to the acoustic modes of single fluid stars, as their dynamics
is mainly dominated by the co-moving degrees of freedom (and they are therefore restored by
the total pressure gradients etcetera). Meanwhile, the superfluid
modes exist only in multi-component stars, and the
counter-phase motion of the perturbed fluid elements dominates
their dynamics. Apart from in particular cases, a neat
decomposition of co- and counter-moving motion into ``ordinary'' and
``superfluid'' modes is not possible. A typical acoustic mode tends to exhibit
\underline{both} degrees of freedom. Nevertheless, in order to distinguish between
these two families of acoustic modes, we denote the ordinary
(superfluid) f- and p-modes with the symbols ${}^l \textrm{f}^{\hspace{0.4mm}\textrm{o}}$ and
${}^l\textrm{p}_{n}^{\textrm{o}}$ (${}^l \textrm{f}^{\hspace{0.4mm}\textrm{s}}$ and ${}^l\textrm{p}_{n}^{\textrm{s}}$),
respectively. 

The spectrum of superfluid neutron stars exhibits 
another important difference with respect to the standard single fluid
models. Superfluid neutron stars do not support an independent class
of g-modes, even if the model has composition
gradients~\citep{1995A&A...303..515L, 2001MNRAS.328.1129A,
  2002A&A...393..949P}.   In a perturbed two-fluid system,
deviations from chemical equilibrium tend to excite the
counter-moving degree of freedom.  The buoyancy only introduces small
corrections to the superfluid acoustic
mode frequencies~\citep[see][for more details]{2001MNRAS.328.1129A}.

\subsection{The polar perturbation problem} \label{sec:polar}

The expansion of the polar perturbations in vector harmonics is given
by the following expressions:
\begin{eqnarray} 
 \xi_{r}^{\x} = W^{\x}(r) Y_{lm} \, , \qquad \xi_{\theta}^{\x} =
 V^{\x}(r) \, \partial_{\theta} Y_{lm} \, ,\qquad \xi_{\phi}^{\x} =
 V^{\x}(r) \, \frac{\partial_{\phi} Y_{lm}}{\sin \theta} \, , \qquad
 \delta \Phi = \delta \hat \Phi (r) Y_{lm} \, .
\end{eqnarray} 
All scalar perturbations are expanded in terms of the spherical
harmonics $Y_{lm}$, c.f., the perturbed gravitational potential.
The problem becomes more tractable if we work with dimensionless
quantities. In the core, we choose to work with the following set of perturbation
variables:
\begin{eqnarray} 
 y_1  \equiv  \frac{W^{\p}}{r}  \qquad 
 y_2  \equiv  \frac{\delta \tilde \mu_\p}{r g}  \qquad
 y_3  \equiv  \frac{W^{\n}}{r}  \qquad 
 y_4  \equiv  \frac{\delta \tilde \mu _\n}{r g}  \qquad 
 y_5  \equiv  \frac{\delta \hat \Phi}{g r}  \qquad 
 y_6  \equiv  \frac{\delta \hat \Phi'}{g}    \label{eq:ydef}
\end{eqnarray} 
where $g$ is the gravitational acceleration. For spherical stars, this
quantity can be related to the chemical potential by the 
equilibrium equation $g=\Phi'=-\tilde \mu'$.

The core is described by a system of six perturbation equations that
we can write (formally) as
\begin{equation}
\frac{d y_{k}}{dr} = Y_{k} \left[ y_i, \sigma^2,b(r) \right] \qquad
i,k=1\dots 6 \, , \label{eq:Y} 
\end{equation}
where the variables $y_{i}$ are defined in~(\ref{eq:ydef}), and the function $b(r)$
denotes a generic background field. The complete system of linearised
equations is given in Appendix~\ref{sec:Core}.

In the crust, the perturbation equations are evidently more complex
due to the presence of the elastic strain tensor.  As in single fluid
models, it is important to choose a set of
perturbation variables that makes the solution of the problem, including the  implementation of the
crust/core junction conditions, easier. In the inner crust, where a mixture of
superfluid free neutrons coexist with a lattice of heavy nuclei, we
define the following set of eight dimensionless perturbation
variables:
\begin{eqnarray} 
z_1 & \equiv & \frac{W^{\ch}}{r}  \qquad 
z_2  \equiv  - \frac{1}{q_\ch} \left( \rho_\ch \Delta \tilde \mu_\ch
  + \frac{2}{3} \mu  \nabla_{i}  \xi_{\ch}^{i} - 2 \mu W'_{\ch} \right) \qquad 
z_3  \equiv  \frac{V^{\ch}}{r}  \qquad
z_4  \equiv \frac{\mu}{q_\ch} \left( V'_\ch - \frac{V_\ch}{r} +
  \frac{W_\ch}{r} \right)  \nn \\
z_5 & \equiv & \frac{W^{\f}}{r}  \qquad 
z_6  \equiv  \frac{\Delta \tilde \mu _\f}{r g}  \qquad 
z_7  \equiv  \frac{\delta \Phi}{r g}  \qquad 
z_8  \equiv  \frac{\delta \Phi'}{g}   \label{eq:zdef}
\end{eqnarray} 
where $q_{\ch}$ has the same units as the shear modulus, and has been
introduced to make $z_2$ and $z_4$ dimensionless. In this work, we
choose $q_{\ch} = p_0$, where $p_0$ is the pressure at the origin
$r=0$. In the function $z_2$, the divergence can be expressed in terms
of the harmonic expansion variables since
\begin{equation}
 \nabla_{i} \xi ^{i}_{\ch} = W'_{\ch} + \frac{2}{r} W_{\ch} -
 \frac{l(l+1)}{r} V_{\ch} \, .  
\end{equation}
The variable $z_2$ and $z_4$ are directly related to the radial and
angular traction components (see Sec~\ref{sec:CCbc} ).  This
 simplifies the implementation of the junction conditions.
Another difference with respect to the functions used in the
core ($y_{k}$) is that $z_2$ and $z_6$ depend on the Lagrangian
perturbation of the chemical potential.  This choice provides a
simpler boundary-value problem and less complex equations in
the core. In the single fluid limit, i.e. when $x_{\ch} = 1$, the
perturbation variables~(\ref{eq:zdef}) reduce to the quantities used
by~\cite{1988ApJ...325..725M} provided we set $q_{\ch} =
p$.\footnote{In \cite{1988ApJ...325..725M}, the definition of the
  variable $z_{2}$ contains a typo. The total mass density $\rho$ must
  be replaced by the pressure $p$.} Naturally, the
variables $z_{5}$ and $z_{6}$ are obsolete  in this limit.

The linearised equations of the crust form a system of eight ordinary
differential equations that we can  write (again, formally) as:
\begin{equation}
\frac{d z_{k}}{dr} = Z_{k} \left[ z_i, \sigma^2, \mu, b(r) \right] \qquad
i,k=1\dots 8 \, . \label{eq:Z} 
\end{equation}
We provide the complete form of these
equations in Appendix~\ref{sec:Crust}.

In the outer crust, all the neutrons are confined to nuclei,
i.e. we have $x_{\f} = 0$. Therefore, the perturbation problem reduces to the
single-fluid case. The oscillation dynamics is then
described by a system of six ordinary differential equations, where
the variables $z_{5}$ and $z_{6}$ are no longer used.

\subsection{Boundary Conditions} \label{sec:polbc}

To solve equations~(\ref{eq:Y}) and~(\ref{eq:Z}) as an eigenvalue
problem, we must impose regularity conditions at the centre, boundary conditions at the
surface of the star, and junction conditions at the two internal
interfaces. The first internal interface marks the transition between
the outer and the inner crust, and the second represents the point where the inner crust gives way to
the fluid core.

At the origin ($r=0$), the regularity conditions lead to the following relations:
\begin{eqnarray}
l \,  y^{0}_{2} & = & \frac{1-\veps_\ch}{g_0} \, \sigma^2  y^{0}_{1} 
                   +  \frac{\veps_\ch}{g_0} \, \sigma^2  y^{0}_{3}
		   -  y^{0}_{6} \, , \\
l \, y^{0}_{4} & = &  \frac{\veps_\n}{g_0} \, \sigma^2  y^{0}_{1}
                   + \frac{1-\veps_\n}{g_0} \, \sigma^2  y^{0}_{3} 
		   -  y^{0}_{6} \, ,\\  
l  \, y^{0}_{5} & = &  y^{0}_{6} \, , 
\end{eqnarray}
where $g_0$ and $y_{k}^{0}$ are, respectively, the gravitational
acceleration and the perturbation variables at the centre.

In Sec.~\ref{sec:CCbc}, we determined the junction conditions at the
crust/core surface $r=R_{cc}$. In terms of the dimensionless
perturbation variables these conditions take the form
\begin{eqnarray}
z_{1} & = & y_{1} \, ,\\
z_{2} & = & \frac{\rho}{q_\ch} r g \left[ x_\ch y_{1} - x_\p y_{2} + \left(x_\p - x_\ch \right) y_4 \right]  \, , \\
z_{4} & = & 0 \, , \\
z_{5} & = & \frac{x_\p - x_\ch }{1-x_\ch} y_{1} + \frac{1-x_\p}{1-x_\ch} y_{3} \, ,\\
z_{6} & = & y_{4} - z_{5} \, , \\
z_{7} & = & y_{5} \, , \\
z_{8} & = & y_{6} \, .
\end{eqnarray}

The boundary conditions that must be satisfied at the inner/outer
crust interface are the continuity of the radial displacement $z_{1}$,
the traction components $z_{2}$ and $z_{4}$, the gravitational
potential perturbation $z_{7}$ and its radial derivative $z_{8}$. The
component of free superfluid neutrons vanishes on this surface. Given this, we
impose a zero boundary condition for the Lagrangian perturbation of
the free neutron chemical potential $(\Delta \tilde \mu_{\f})$,
i.e. take $z_{6} = 0$.

Beyond the transition to the outer crust, the problem is equivalent to
the single fluid approximation. Therefore, at the stellar surface we
must impose the continuity of the traction perturbation and a
condition for the gravitational perturbation.  The $r \to R$ limit of
the linearised Poisson equation leads to the following
expression~\citep{1989nos..book.....U, 2002A&A...393..949P}:
\begin{equation}
\delta \Phi ' + \frac{l+1}{R} \delta \Phi = - \xi^r \lim_{r\to R^{-}}
\rho(r) \, , 
\end{equation}
For a model such that the density vanishes at the surface, this means that 
we should have
\begin{equation}
\delta \Phi ' + \frac{l+1}{R} \delta \Phi = 0 \, . 
\end{equation}
At the star's surface, the perturbation variables must then satisfy the
following relations:
\begin{eqnarray}
z_{1} & = & 1 \, \\
z_{2} & = & 0 \, \\
z_{4} & = & 0 \, \\
z_{8} & = & - ( l+1) z_{7} \,  ,
\end{eqnarray}
where we have imposed a normalization condition for the confined
nucleon Lagrangian displacement, $z_{1} = 1$. 

The outer crust of the stellar models studied in this work is very
thin. Neutrons start to drip out from the nuclei at $r=R_{\rm{ND}}
\cong 0.9988 R$. The mode frequency calculation is not really affected if we
impose the boundary condition $\Delta
\tilde \mu_{\f} = 0$ ($z_{6} = 0$) relevant to the inner/outer crust interface
at the surface of the star. This is not surprising. The fact that the low density region has little effect on the 
global oscillations of the star also means that we can ignore the thin ocean that should exist below the density at which the 
crust is expected to melt (at finite temperatures).

\subsection{The axial perturbation problem}

Axial perturbations of spherical stars describe torsional
oscillations which are sustained by the crust's shear. The corresponding vector
harmonic expansion of the Lagrangian displacement is given by 
\begin{eqnarray} 
 \xi_{r}^{\x}      =  0 \, \qquad
 \xi_{\theta}^{\x}  =  U^{\x}(r) \, \frac{\partial_{\phi} Y_{lm}}{\sin \theta} \, , \qquad 
 \xi_{\phi}^{\x}   = -  U^{\x}(r) \, \partial_{\theta} Y_{lm} \,
 ,  \qquad \label{eq:ax}
\end{eqnarray} 
while all Eulerian perturbations of the scalar variables are
zero. From equation~(\ref{eq:ax}), it is clear that the vector
$\xi_{\x}$ has zero divergence, $\nabla_{i} \xi_{\x}^{i} =
0$. Therefore, we have $\Delta \rho_\x = \Delta \tilde \mu _\x =
0$.

Substitution of the expansion~(\ref{eq:ax}) into the linearised
equations~(\ref{eq:eul-3})-(\ref{eq:eul-4}) leads to a system of two
equations:
\begin{eqnarray}
\sigma ^{2} \veps_{\star}^{-1}  U_{\ch} & = &  
\frac{\mu'}{\rho_\ch} \left(\frac{U_\ch}{r} -  U_{\ch}'  \right) 
                                   -    \frac{\mu}{\rho_\ch} \left[
                                    U_{\ch}'' + \frac{2}{r} U_\ch' -
                                    \frac{l(l+1)}{r^2} U_\ch \right]
                                  \, , \label{eq:Uc} \\ 
U_{\f} & = &  - \frac{\veps_{\f}}{1-\veps_{\f}}  U_{\ch}
\,   . \label{eq:Uf} 
\end{eqnarray}
where we have defined 
\begin{equation}
\veps_{\star} \equiv \frac{ 1 - \veps_{\n} }{1-\bar \veps} = 
\frac{ 1 - x_\ch \bar \veps }{1-\bar \veps} \,  .  \label{eq:epsstar}
\end{equation} 
As before, we have assumed a harmonic time dependence with frequency $\sigma$, as before. The system
of equations~(\ref{eq:Uc})-(\ref{eq:Uf}) describes the linear dynamics
of torsional oscillations in the inner crust. In the outer crust,
there are no free superfluid neutrons, which means that we have $x_{\f} = 0$. Therefore,
equation~(\ref{eq:Uf}) becomes obsolete, while equation~(\ref{eq:Uc})
reduces to the single fluid approximation  provided $x_{\ch}
=1$ and the entrainment is set to zero, i.e. $\veps_{\x} = 0$.

It is worth noting that, in both the outer and inner crust, equation~(\ref{eq:Uc}) depends only
on the $U_{\ch}$ variable, which therefore completely determines the torsional
oscillation modes. That this is the case also in General Relativity has already been demonstrated by  \cite{2009CQGra..26o5016S}.
Once the eigenvalue problem is solved for the confined nucleons, we
can determine the eigenfunctions for the free neutrons in the inner
crust from equation~(\ref{eq:Uf}).

As in the polar case, we introduce a new system of dimensionless
variables in order to simplify the boundary value problem. These variables are
defined as:
\begin{eqnarray} 
 s_1  \equiv  \frac{U^{\ch}}{r} \, ,  \qquad 
 s_2  \equiv  \frac{\mu}{q_{\ch}} \left( U_{\ch}' - \frac{U^\ch}{r} \right) \,  ,
\end{eqnarray}
where $s_2$ is proportional to the transverse component of the
traction.  Equation~(\ref{eq:Uc}) now becomes a system of two first
order ODE:
\begin{eqnarray}
r \frac{d  s_1}{dr} & = & \frac{q_{\ch}}{\mu}   s_2\, , \label{eq:ax1} \\
r \frac{d  s_2}{dr} & = & \left[ \mu \left(l+2\right)\left(l-1\right)
  - \frac{\rho_\ch}{\veps_\star} r^2 \sigma^2 \right] \frac{s_1}{q_{\ch}} - 3 s_2
\, ,  \label{eq:ax2} 
\end{eqnarray}
The relevant boundary conditions are; the continuity of the traction at the
crust/core interface, inner/outer crust interface and at the stellar
surface. We therefore impose $s_2 = 0$ both at the crust/core
boundary and at the star's surface. In addition, we normalize the
radial displacement by setting $s_1=1$ at the stellar surface.

\begin{figure}
\begin{center}
\includegraphics[height=70mm]{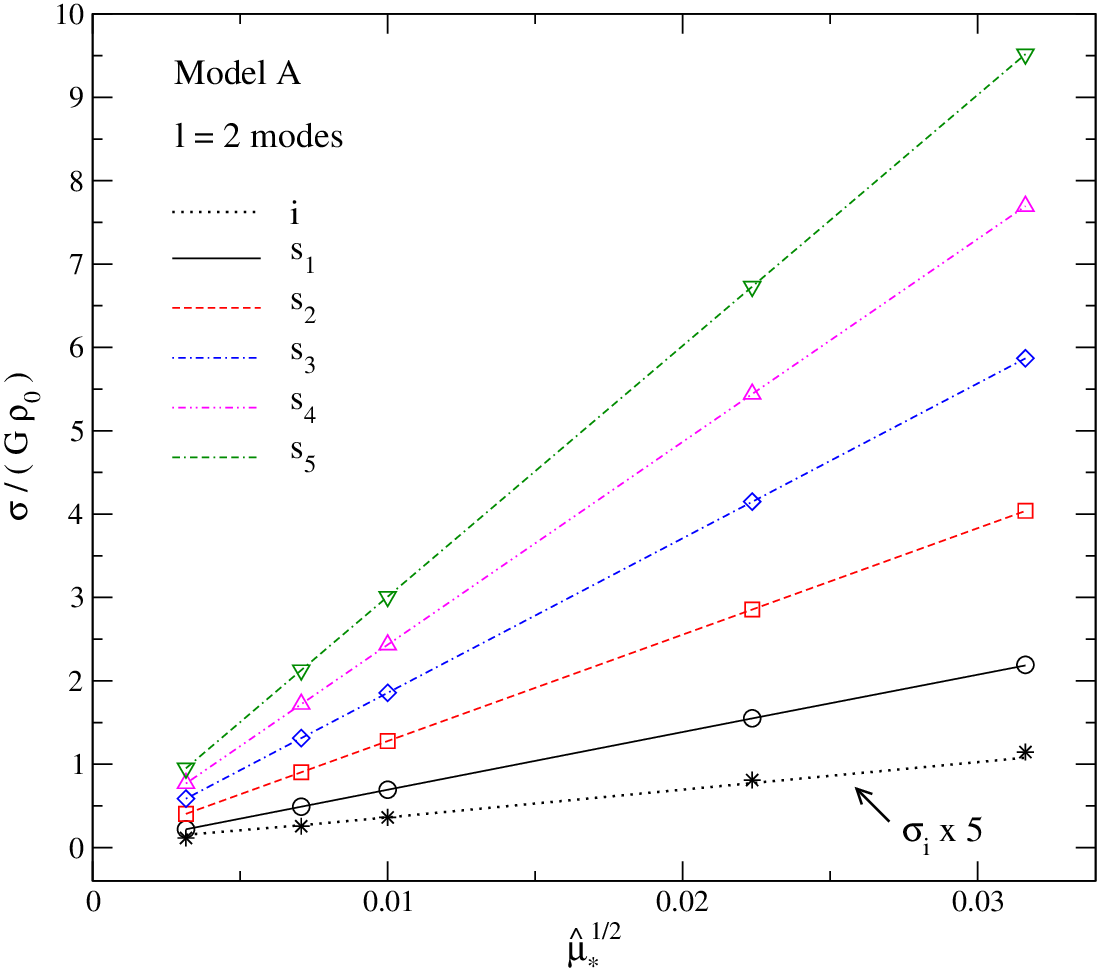}
\includegraphics[height=70mm]{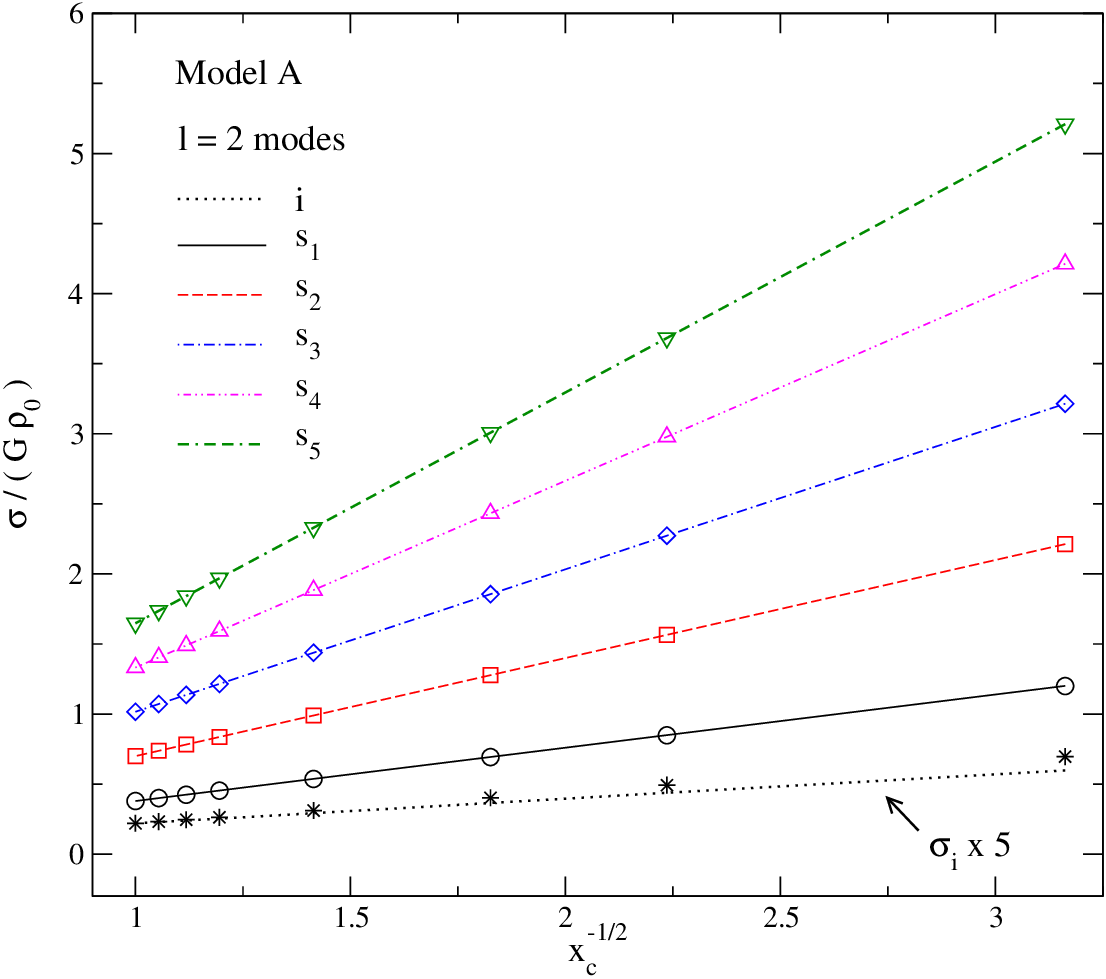}
\caption{ For non-stratified models A with zero entrainment, this
  figure displays the $l=2$ interface mode and the first five $l=2$
  spheroidal shear modes.  In both panels, the numerical results are
  shown as lines, while the values estimated from the plane-wave
  equation~(\ref{eq:sh}) are represented by symbols. In the left panel, we show the
  dependence of the interface and shear mode frequencies (vertical
  axis) on the elastic modulus parameter $\hat \mu _{\ast}$
  (horizontal axis) for a model A with constant $x_{\p} = 0.1$ and
  $x_{\ch} = 0.3$.  In the right panel, we study the dependence of the
  spectrum on the confined component fraction $x_{\ch}$ (horizontal
  axis). In this case, the models A have $\hat \mu _{\ast} =
  10^{-4}$. In both cases, equation~(\ref{eq:sh}) describes very well
  the spectrum of the spheroidal shear modes.
  \label{fig:3}}
\end{center}
\end{figure}

\section{Results}

Let us now consider the detailed oscillations of the multi-component neutron star model
developed in the previous sections. We focus on non-rotating,
polytropic two-fluid models, and study the effects of varying  component
fractions,  crust elasticity and entrainment on both the shear modes and acoustic
waves.  Building on the discussion in  Sec.~\ref{sec:EP},  we decompose the problem
into polar and axial
perturbations. The spectrum of each family of perturbations can be
studied as a boundary value problem by solving the system of
equations~(\ref{eq:Y}) and~(\ref{eq:Z}) for the polar modes
and~(\ref{eq:ax1})-(\ref{eq:ax2}) for the axial modes. In both cases 
we use a shooting method to a fixed point. For the polar sector, we
set the fixed point at the crust/core interface, where the core and
crust solutions are matched using the junctions conditions discussed in
Sec.~\ref{sec:CCbc} and~\ref{sec:polbc}.

\subsection{Shear modes}

In simple plane-wave models,  the frequency of the crust shear waves is proportional to the shear velocity;
\begin{equation}
  v_{s} = \sqrt{\check \mu / \rho_{\ch}} \,  .  \label{eq:vs}
\end{equation}
However, in the case of a 
two-fluid model the shear velocity depends on the
mass density of the confined nucleons and not on the total mass
density, as in the single fluid 
case~\citep{2009MNRAS.396..894A}.  Still, considering typical parameters, it is easy to see that the shear waves have low
frequencies. One would also expect the shear model to be predominantly transverse. They should be 
present in both the polar and the axial sector, and we will refer to the two classes as 
spheroidal and torsional shear modes, respectively. The crust elasticity will also
affect the acoustic mode frequencies by introducing very small
corrections (roughly of the order of the ratio between the shear wave speed and the speed of sound)
to the acoustic modes.  Furthermore, the free
superfluid neutrons in the inner crust may interact with the lattice
of nuclei through entrainment and  modify
the spectrum. The main impact of the entrainment is expected in the
deep crust, where the effective mass of the free neutrons may be very
large~\citep{2005NuPhA.747..109C}, c.f., Figure~\ref{fig:bm2}.

From a plane-wave analysis of the two-fluid problem, we can derive the frequency
dependence of the shear modes on the mass density, the entrainment and the
shear modulus~\citep{2009MNRAS.396..894A}:
\begin{equation}
\sigma = \sqrt{ \frac{\veps_{\star} \check \mu }{\rho_\ch}} k \, ,  \label{eq:sh}
\end{equation} 
where $k$ is the magnitude of the wave-vector. We are interested in testing the accuracy of
this rough approximation against full
numerical results for stratified and non-stratified superfluid models.

The shear modulus of the crust can be determined from~\citep{1991ApJ...375..679S}:
\begin{equation}
\check   \mu = \frac{0.1194}{1+0.595 \left( \Gamma_0 / \Gamma \right)^2 } 
  \frac{ n_i \left( Z e \right)^2 }{a} \label{eq:mod} \,  .
\end{equation}
This result was inferred from Monte Carlo simulations of the Coulomb interactions in the neutron
star crust. The quantity $n_i$ is the ion
density, $Z$ the atomic number of the nuclei and $a=\left( 3/4\pi
  n_{i} \right) ^{1/3}$ is the average inter-ion spacing. The ratio
between the Coulomb and thermal energy is parameterised by $\Gamma =
\left( Z e \right)^2 / a k_{B} T$, where $T$ is the temperature and
$k_{B}$ the Boltzmann constant. \cite{1993PhRvE..47.4330F} have discussed
plausible values for the parameter $\Gamma_{0}$.

To be truly consistent with~(\ref{eq:mod}), a stellar model 
should obviously be obtained from a realistic
EoS. As we have already discussed,  we adopt a different strategy in this work. We focus on simple
polytropic two-fluid models and determine the shear modulus from a phenomenological 
relation. Detailed calculations for realistic EoSs (see~\cite{2001A&A...380..151D}) suggest that the \underline{specific} shear modulus  is almost constant across the
crust, i.e.
\begin{equation}
\frac{\check \mu }{ \rho} \simeq 10^{16}  \textrm{cm}^{2} \textrm{s}^{-2} \,
. \label{eq:spmu}
\end{equation}
Hence, it is reasonable to (as a first approximation) consider simple models such that 
\begin{equation}
\check \mu = \mu_{\star} \rho \, ,
\end{equation}
where $\mu_{\star}$ is a constant parameter. 
We consider only this simple model, and 
in the numerical code, we use its dimensionless form $\hat
\mu_{\ast}$, which is defined as follows:
\begin{equation}
\mu_{\ast} = \hat \mu_{\ast}  \, G \rho_0 R^2 \, . \label{eq:hmu}
\end{equation}
The constant parameter~(\ref{eq:hmu}) may be estimated at the
crust/core interface by using equation~(\ref{eq:spmu}), the
transition mass density $\rho_{cc}=1.2845\times10^{14}~\text{g
  cm}^{-3} $ at $r=R_{cc}$, and assuming typical neutron star values
for the central mass density $\rho_{0} = 10^{15}$~g~cm$^{-3}$ and
radius $R=10$~km. This way we obtain $\hat \mu_{\ast} \simeq 10^{-4}$ at
$r=R_{cc}$.
\begin{figure}
\begin{center}
\includegraphics[height=70mm]{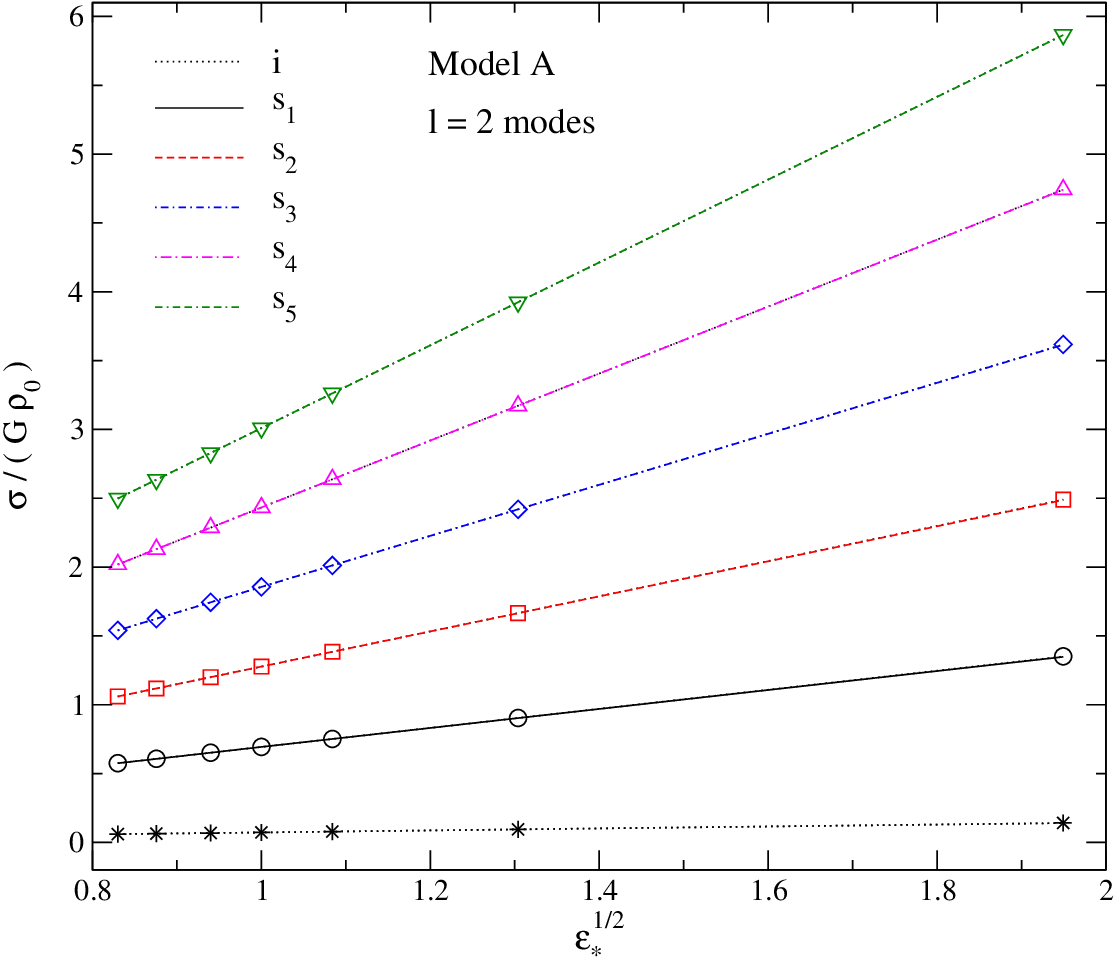}
\caption{ For a model A with constant $x_{\p}
  = 0.1$, $x_{\ch} = 0.3$, and $\hat \mu
  _{\ast} = 10^{-4}$, we show in this figure the frequency dependence of the
  $l=2$ interface and spheroidal shear modes on the entrainment
  parameter $\veps_{\star}$ (horizontal axis). The numerical and
  estimated frequencies are represented using the same notation as
  in Fig.~\ref{fig:3}. The results show that the shear mode
  eigenfrequencies are well approximated by equation~(\ref{eq:sh}). 
  \label{fig:4}}
\end{center}
\end{figure}

\subsubsection{Spheroidal shear and interface
  modes} \label{sec:spsh}

We consider first the spheroidal shear modes of the set of A models;
stellar configurations with constant component fractions. In order to explore the 
dependence on the shear modulus, we focus on a model with $x_\p = 0.1$, 
$x_{\ch} = 0.3$ and zero entrainment. The resulting frequencies of the $l=2$
shear modes  and the interface mode are shown in the left
panel of Fig.~\ref{fig:3} for different values of the shear
modulus ($\hat{\mu}_*$). We compare our numerical results to the plane-wave
relation~(\ref{eq:sh}). To do this we first ``infer'' the effective value of the wave vector $k$ in equation~(\ref{eq:sh})
for a single
numerical data point $(\sigma,\check \mu)$ and then use this value in
equation~(\ref{eq:sh}) to extend the relation to other parameter values.   
From the results in Fig.~\ref{fig:3},  we see that the numerical results scale according to the 
phenomenological relation. The agreement is, in fact, very good for the shear
modes. Meanwhile, the interface mode frequencies deviate from the
plane-wave relation. This is expected since  these modes have a different
origin; they depend sensitively on the physical conditions near the
crust/core interface.  Interface modes typically have small frequencies and amplitudes concentrated  
around the internal interfaces. A characteristic property of these modes is a cusp in the radial displacement 
eigenfunction at the internal boundary layer.  In this work, we do not consider neutron stars with an ocean, which generates 
another interface mode associated with the ocean/crust transition 
\citep[see][for more details]{1988ApJ...325..725M, 2005ApJ...619.1054P}.

\begin{figure}
\begin{center}
\includegraphics[height=70mm]{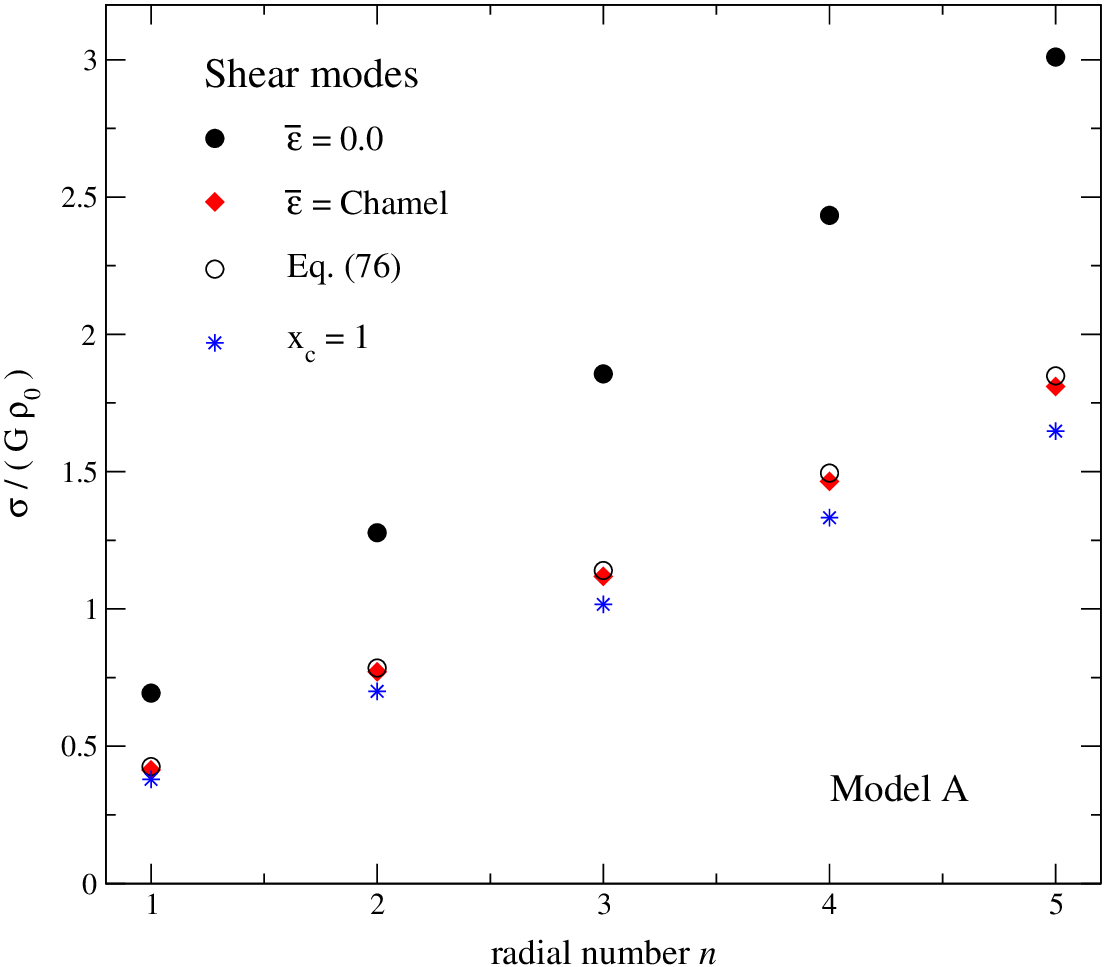}
\includegraphics[height=70mm]{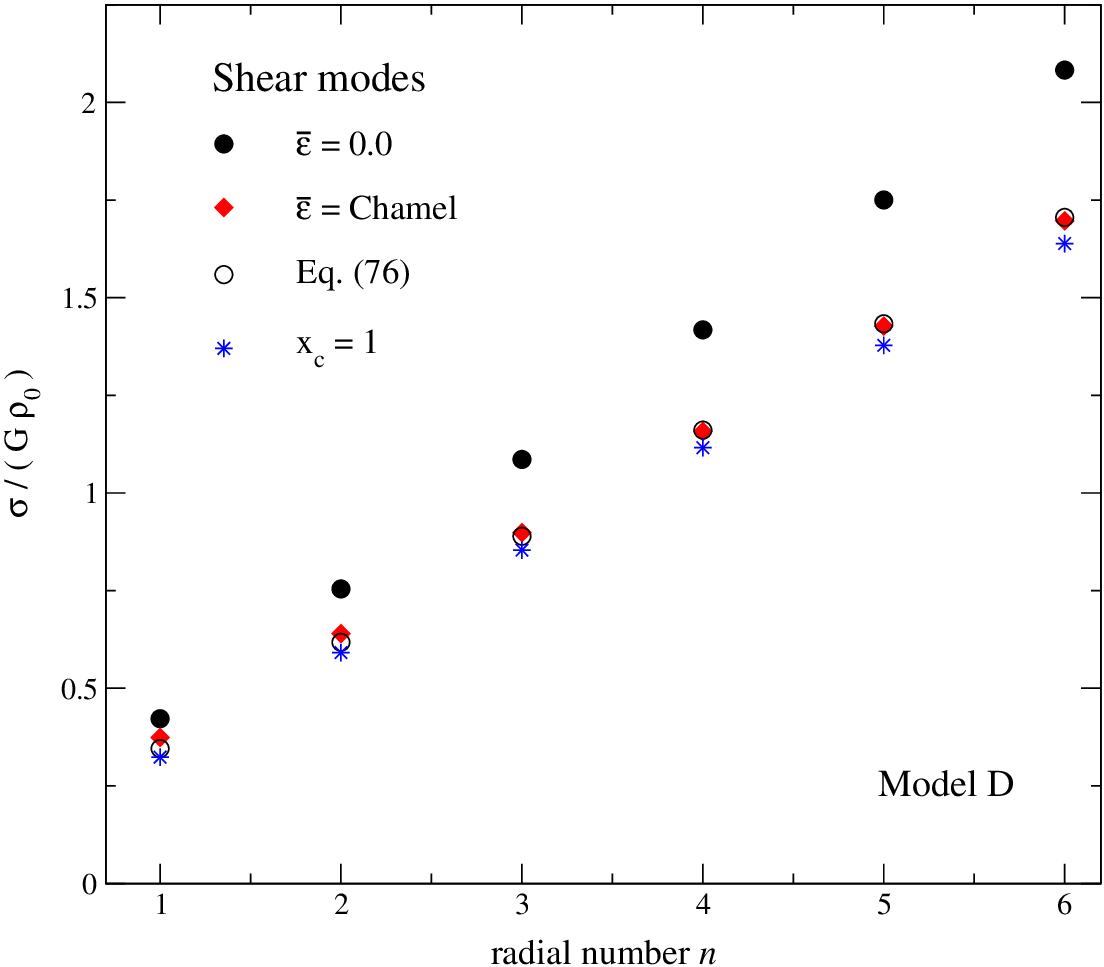}
\caption{ This figure displays the $l=2$ spheroidal shear mode eigenfrequencies for
  model A (left panel) and D (right panel). In both cases, the shear
  modulus parameter is $\hat \mu _{\ast} = 10^{-4}$. On the horizontal axis, we
  show the radial number $n$ of the shear mode ${}^2 \textrm{s}_{n}$, while the vertical
  axis displays the dimensionless mode frequencies. The frequencies for 
  the zero-entrainment model are shown as filled circles, those
  corresponding to the Chamel entrainment as filled diamonds, while
  the open circles are the values estimated from equation~(\ref{eq:sh3}). 
  We show also the shear mode frequencies (denoted by stars) for models \underline{without}  
  free  superfluid neutrons in the crust  
  ($x_{\ch}=1$). 
  \label{fig:5}}
\end{center}
\end{figure}

Let us now apply the same strategy to the dependence on 
the fraction of confined nucleons $x_{\ch}$. In this
case, we consider a model A with $x_\p = 0.1$, zero entrainment and
$\hat \mu_{\ast} = 10^{-4}$. The results shown in the right panel of
Fig.~\ref{fig:3} confirm the good agreement between the numerical
shear-mode frequencies and the analytical relation~(\ref{eq:sh}) also in this case.

We can also study the dependence on the entrainment parameter
$\veps_{\star}$. To do this we construct Model A to be a star with fractions $x_\p = 0.1$,
$x_{\ch} = 0.3$, and shear modulus $\hat \mu_{\ast} = 10^{-4}$.  We
consider two distinct entrainment configurations. In the first case,
we set the parameter $\bar \veps$ constant and determine $\veps_\x$
from the relation $\veps_\x = \bar \veps / x_{\y} $. 
We have already considered this configuration in a different context~\citep{2009MNRAS.396..951P, 2011MNRAS.413...47P}. In the second
case, we investigate the entrainment model described in
Sec.~\ref{sec:entr}, which is based on the work
of~\citet{2005NuPhA.747..109C, 2006NuPhA.773..263C,
  2008MNRAS.388..737C}.
When $\bar \veps$ is taken to be constant, equation~(\ref{eq:sh}) describes
the frequencies of the shear modes very well, see figure~\ref{fig:4}. However, the
plane-wave approximation is not accurate for the
model with a realistic entrainment profile. This result is (obviously) expected, as the plane-wave approximation follows from a local analysis
(based on constant parameters). Nevertheless,
we find that an ``average'' of equation~(\ref{eq:sh}) over 
the crust reproduces the shear frequencies of this
model rather well. This means that we can use the following shear mode relation:
\begin{equation}
\sigma^2 =  \frac{ \langle \veps_{\star}  \rangle_{av} }{ \langle \rho_\ch \rangle_{av} } \, \mu \,  k^2 \, ,  \label{eq:shav}
\end{equation} 
where $\langle ~ \rangle _{av}$ denotes an average over the
crust. In the left panel of Fig.~\ref{fig:5}, we compare the shear
mode frequencies of model A without entrainment, i.e.
with $\bar \veps = 0$, to a model based on ``realistic'' entrainment. The results show that the
relation between these two sets of data is accurately
described by equation~(\ref{eq:shav}), i.e.,
\begin{equation}
\sigma \left( \veps_{\star}  \right)  = \sqrt{ \langle \veps_{\star}
  \rangle _{av} } \sigma_0 \,  , \label{eq:sh3}
\end{equation}
where $\sigma_0$ is the frequency for the zero entrainment model. 

Next we consider the effect of composition gradients on the
shear modes, making use of the D models introduced in Sec.~\ref{sec:EoS}.  With two distinct values for the polytropic indices of the core
and the crust, these models mimic the composition of
realistic EoS, i.e., the proton fraction decreases from a central value, while the
 fraction of confined nucleons in the crust increases away from the interface.  We study a
sequence of models with the same core configuration, such that
$x_\p=0.2$ at $r=0$, and different constituent fractions in the crust.
More details on these models D are given in Sec.~\ref{sec:EoS}.
Before quantifying the effect that a composition gradient has on the shear
modes, we consider the realistic entrainment profile for these
stratified models.  To this end, we select a model D with $x_\ch =
0.3$ at the crust/core interface and let $\hat \mu_{\star} =
10^{-4}$. The obtained shear mode eigenfrequencies are shown in the right panel
of Fig.~\ref{fig:5}. Approximate frequencies determined from the
formula~(\ref{eq:sh3}), are (again) in good agreement with the numerical
values.

In order to emphasize the importance of 
entrainment on the spectrum, we also show in Fig.~\ref{fig:5} the shear mode frequencies for stellar models without 
free superfluid neutrons in the crust ($x_{\ch}=1$).  With a large effective mass, the relative motion between the free superfluid neutrons 
and the crust is restricted. Therefore, the shear mode frequencies tend to the values expected for models without a superfluid component.   
This effect is clear when we consider the plane-wave approximation~(\ref{eq:sh}) in the limit $m^{\star}_{\n} \gg m$.
In this case, we have $\veps_{\star} \approx x_{\ch}$ and equation~(\ref{eq:sh}) becomes~\citep{2009MNRAS.396..894A}
\begin{equation}
\sigma \approx \sqrt{ \frac{\check \mu }{\rho}} k \, ,  \label{eq:shS}
\end{equation}
which is the formula expected for pure elastic crust without superfluid neutrons~\citep{1988ApJ...325..725M}. For the stellar models 
and effective masses considered in this work, there is roughly a 10\% difference between the shear mode spectrum determined with and without 
a superfluid component in the crust.

\begin{figure}
\begin{center}
\includegraphics[height=70mm]{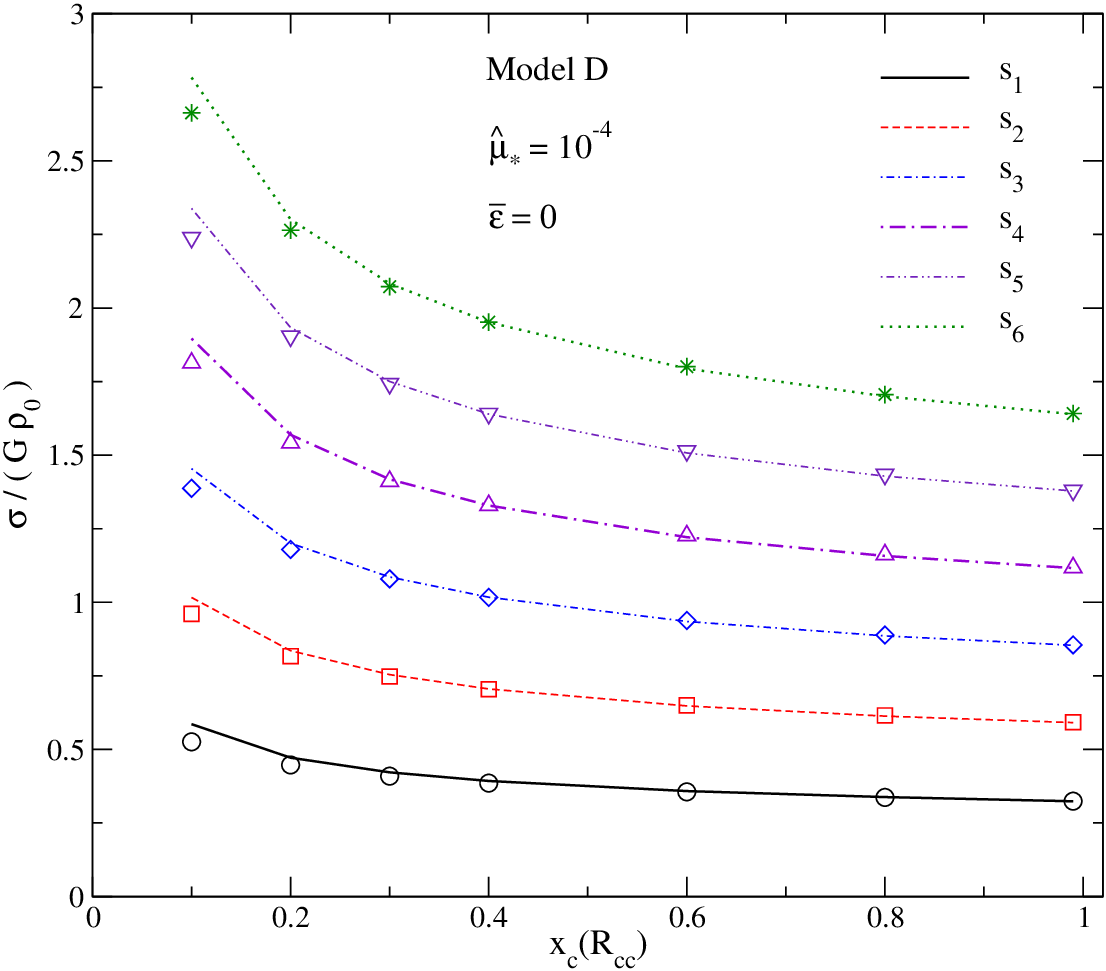}
\includegraphics[height=70mm]{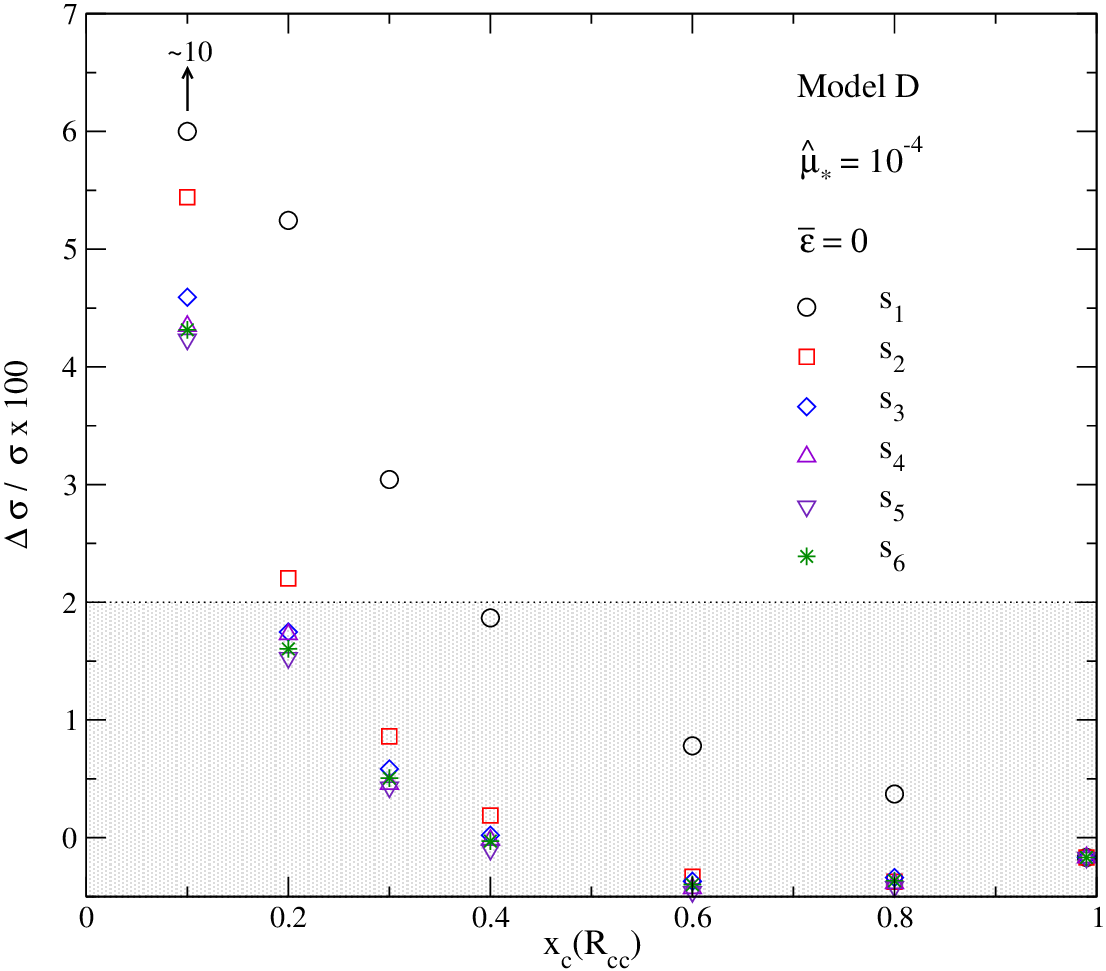}
\caption{This figure shows the effects of composition gradients on the
  spheroidal shear modes. The background stars are described by models
  D with different $x_{\ch}$ at the crust/core interface, $r=R_{cc}$,
  but with the same shear modulus, $\hat \mu _{\ast} = 10^{-4}$, and
  zero entrainment. In the left panel, we show the dimensionless
  eigenfrequencies of the first six $l=2$ spheroidal shear modes
  (lines) and their values estimated with equation~(\ref{eq:sh4})
  (symbols).  The accuracy of this equation is shown in the right
  panel, where we report the relative error between the code results
  and the simple analytical relation~(\ref{eq:sh4}). In both panels,
  the horizontal axis displays the confined nucleons fraction
  $x_{\ch}$ at $r=R_{cc}$. A horizontal dotted line marks the 2\%
  error threshold in the right panel.  \label{fig:6}}
\end{center}
\end{figure}

Finally, we study the effects of composition stratification on the
spectrum. We consider a sequence of models D, where we set $\hat
\mu_{\star} = 10^{-4}$ and take the entrainment to vanish. The original plane-wave
relation~(\ref{eq:sh}) is (obviously) not adapted to describe the shear
modes in stratified stars. However, we obtain a useful 
description of the spectrum from equation~(\ref{eq:shav}), according to 
which we have;
\begin{equation}
\sigma  \left( x_{\ch} \right) = \frac{ \sigma_{1} }{ \sqrt{ \langle x_{\ch} \rangle_{av} } } \,  , \label{eq:sh4}
\end{equation}
where $\sigma_1$ is the frequency for the non-stratified, $\langle x_{\ch} \rangle _{av}= 1$,
case. Figure~\ref{fig:6} shows the frequencies of the first six shear
modes for the set of D models. In the left panel, the numerical eigenfrequencies 
are compared to the values estimated from
equation~(\ref{eq:sh4}). The accuracy of the averaged
plane-wave relation decreases when $x_{\ch}$ becomes smaller at the
crust/core interface. This behaviour is better illustrated in the
right panel of Fig.~\ref{fig:6}, where we show the relative error of
the approximation. In particular, we note that
equation~(\ref{eq:shav}) provides more accurate results for the realistic
entrainment configuration effects than for the composition
gradients. Therefore, when both the component fraction and the
entrainment vary, the shear modes can be described by
equation~(\ref{eq:shav}) with accuracy limited by the
composition gradient effects.

Tables~\ref{tab:2} and~\ref{tab:3} provide the eigenfrequencies
of the $l=2$ interface mode and the first three $l=2$ spheroidal
shear modes for some selected models A and D (see the table captions
for details). In Table~\ref{tab:2}, the mode frequencies are given for
different confined nucleon fractions $x_{\ch}$. In Table~\ref{tab:3},
the stellar models include the realistic entrainment profile. 
In order to test these results, we can convert the
dimensionless eigenfrequencies of Table~\ref{tab:2} into physical units
and compare them to results from literature. 
As far as we know, there are no published results for shear and interface modes in single fluid
stars with polytropic EoS and Newtonian gravity.  However, there have been a number of
analyses of the shear and interface mode spectrum for
various other neutron star models. We can asses to what extent our
results are consistent with these studies. To this end, we select from the
stellar models A and D described in Sec.~\ref{sec:EoS} the model with
$x_{\ch} = 1$, as the results in the literature rely on the single
fluid ``approximation''. From Table~\ref{tab:2} and the central mass
densities calculated in Sec.~\ref{sec:EoS}, the oscillation
frequencies of the ${}^2 \rm{s}_{1}$ and ${}^2 \rm{s}_{2}$ modes are,
respectively, $\nu=534.68~\textrm{Hz}$ and $\nu=986.49~\textrm{Hz}$
for model A, and $\nu=444.69~\textrm{Hz}$ and $\nu=812.43~\textrm{Hz}$
for model D. For the $l=2$ interface mode, we find
$\nu=62.03~\textrm{Hz}$ and $\nu=83.41~\textrm{Hz}$ for models A and
D, respectively.  These mode frequencies are consistent with the
results obtained by~\citet{1988ApJ...325..725M}. 
%
\begin{table}
\begin{center}
  \caption{\label{tab:2} Frequencies of the $l=2$ interface mode and
    the first three $l=2$ spheroidal and toroidal shear modes for some
    of the stellar models studied in this work.  Frequencies are given
    in units of $\sigma / \sqrt{G \rho_0}$, where $G$ is the
    gravitational constant and $\rho_0$ represents the central mass
    density.  The first column labels the stellar model. The
    non-stratified models A have constant proton fraction, $x_\p =
    0.1$ and the confined nucleon fraction $x_{\ch}$ shown in the
    second column. The stratified models D have the same central
    proton fraction, $x_\p = 0.2$ at $r=0$, while the values of
    $x_{\ch}$ at $r=R_{cc}$ are given in the second column. All models have zero entrainment and $\hat
    \mu_{\star} = 10^{-4}$.  }
\begin{tabular}{ c  c  c c c c c  c c  }
  \hline
  Model &   $ x_{\ch} $  &  $ {}^2 \rm{i} $ & ${}^2 \rm{s}_{1}$  & ${}^2 \rm{s}_{2}$ &
  ${}^2 \rm{s}_{3}$ & ${}^2 \rm{t}_{0}$ & ${}^2 \rm{t}_{1}$ & ${}^2 \rm{t}_{2}$   \\
  \hline
  A  &  0.1   &  0.1195   &  1.2008  &  2.2128  &  3.2143   & 0.0677  & 1.1977 & 2.2041\\
  A  &  0.5   &  0.0585   &  0.4248  &  0.7824  &  1.4377   & 0.0303  & 0.5356 & 0.9857\\
  A  &  1.0   &  0.0440   &  0.3793  &  0.6998  &  1.0165   &  0.0215 & 0.3807 & 0.7005 \\
  \\
 D  &  0.1   &  0.0796    &  0.5858  & 1.0160  &  1.4539  &  0.0525   & 0.5846 & 1.0138 \\
 D  &  0.5   &  0.0589    &  0.3729  & 0.6722  &  0.9697  &  0.0271   & 0.3723 & 0.6718 \\
 D  &  1.0   &  0.0607    &  0.3236  & 0.5912  &  0.8537  &  0.0219   & 0.3256 & 0.5911\\
\hline
\end{tabular}
\end{center}
\end{table}

\begin{table}
\begin{center}
  \caption{This table provides the frequencies of the $l=2$
    interface mode and the first three spheroidal and toroidal shear
    modes for models A and D with the ``realistic'' entrainment
    determined
    by~\citet{2005NuPhA.748..675C,2005NuPhA.747..109C,2006NuPhA.773..263C}.
    See Sec.~\ref{sec:entr} for more details. Frequencies are given in
    dimensionless units, $\sigma / \sqrt{G \rho_0}$. The two stellar
    configurations selected in this table are: i) a model A with
    constant $x_{\p} = 0.1$ and $x_{\ch}=0.3$, and ii) a model D with
    $x_{\p} = 0.2$ at $r=0$ and $x_{\ch}=0.3$ at $r=R_{cc}$. All models have $\hat \mu_{\star} =
    10^{-4}$. \label{tab:3} }
\begin{tabular}{ c  c c c c c  c c  }
  \hline
  Model &  $ {}^2 \rm{i} $ & ${}^2 \rm{s}_{1}$  & ${}^2 \rm{s}_{2}$ &
  ${}^2 \rm{s}_{3}$ & ${}^2 \rm{t}_{0}$ & ${}^2 \rm{t}_{1}$ & ${}^2 \rm{t}_{2}$   \\
  \hline
  A    &  0.0431  &  0.4265 &  0.7765  & 1.1227  & 0.0391  & 0.6915  &  1.2726 \\
  D    &  0.0422  &  0.3793 &  0.6305  &  0.8941 & 0.0238  & 0.3556  &  0.6226 \\
\hline
\end{tabular}
\end{center}
\end{table}

\subsubsection{Torsional modes}  \label{sec:tosh}

Motivated by magnetar seismology, the 
spectrum of toroidal shear modes has already been studied in
relativistic two-fluid neutron stars~\citep{2009CQGra..26o5016S}. Our
Newtonian analysis may therefore seem somewhat superfluous. However, if we want to
improve our current stellar models by considering the effects of
rotation and magnetic field, the perturbation problem becomes more
complex (as the dynamics couples the polar and the axial 
perturbations). As the problem is more tractable in the context of Newtonian gravity, 
it makes sense to provide the corresponding non-rotating results here.

Following the  strategy adopted for the
spheroidal shear modes, we study the dependence of the $l=2$ toroidal
modes on the shear modulus, composition fraction and entrainment.
As previously, the plane-wave relation~(\ref{eq:sh}) describes the
spectrum of the non-stratified models A accurately, while  the averaged
version~(\ref{eq:shav}) is more appropriate for both models D and
models with realistic entrainment. It is not surprising that
the main conclusions drawn for the s-modes remain valid also for the
t-modes, as both sets of oscillations are restored by the crust
elasticity. Given the similarities to the s-modes, we do not illustrate the torsional mode
properties further, but the frequencies of the first three
$l=2$ t-modes are given in Tables~\ref{tab:2} and~\ref{tab:3}.

The frequency of the fundamental torsional mode, $^2 \rm{t}_0$, is typically between 
$10$  and  $50~\textrm{Hz}$~\citep{1988ApJ...325..725M, 2009CQGra..26o5016S, PhysRevLett.103.181101}, 
depending on the chosen model  for the crust.
Using the stellar parameters in Table~\ref{tab:2}, we find that $\nu = 30.31~\textrm{Hz}$ for
model A and $30.09~\textrm{Hz}$ for model D. These values depend on the shear modulus as 
$\hat \mu_{\star} ^{1/2}$, and hence the frequencies assume lower values for $\hat \mu_{\star} < 10^{-4}$.
The results  are certainly consistent with the frequencies expected for the fundamental torsional modes.

\begin{figure}
\begin{center}
\includegraphics[height=70mm]{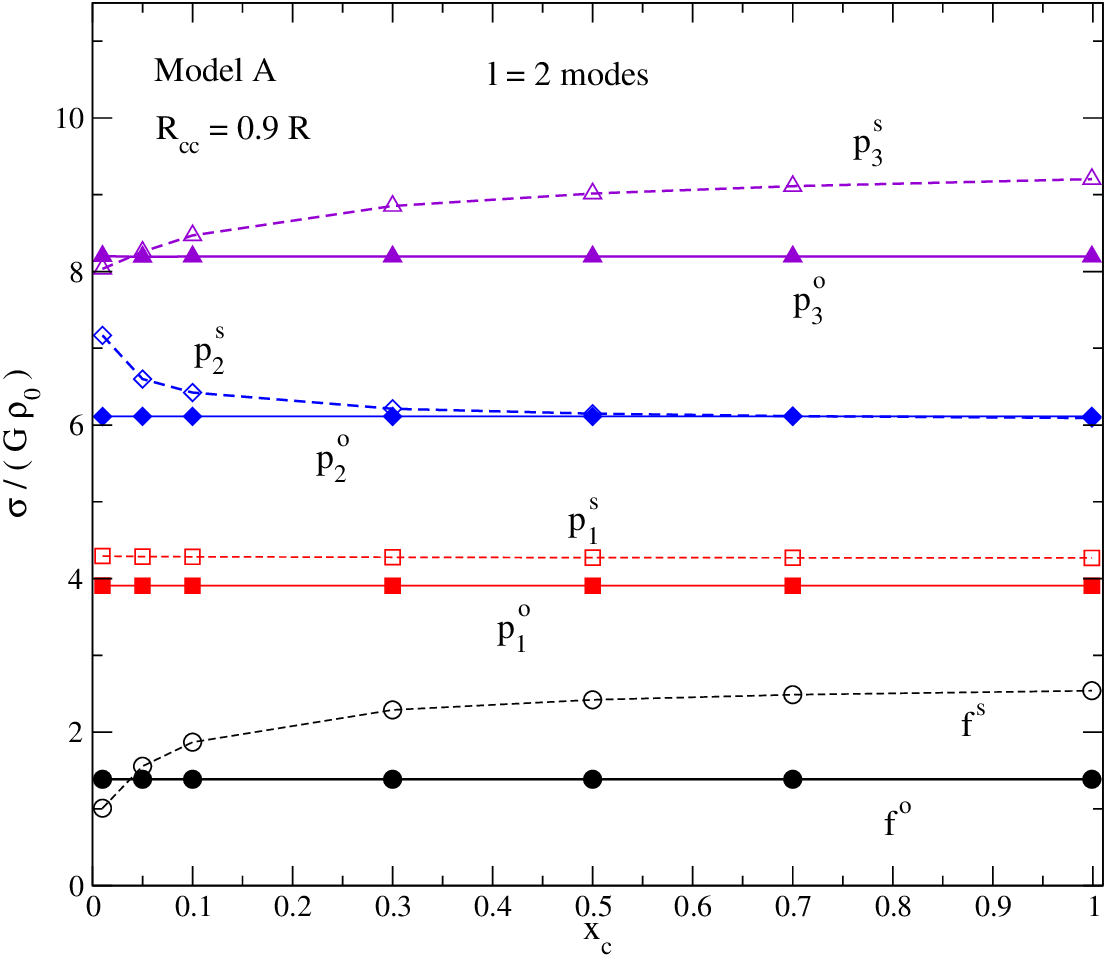}
\includegraphics[height=70mm]{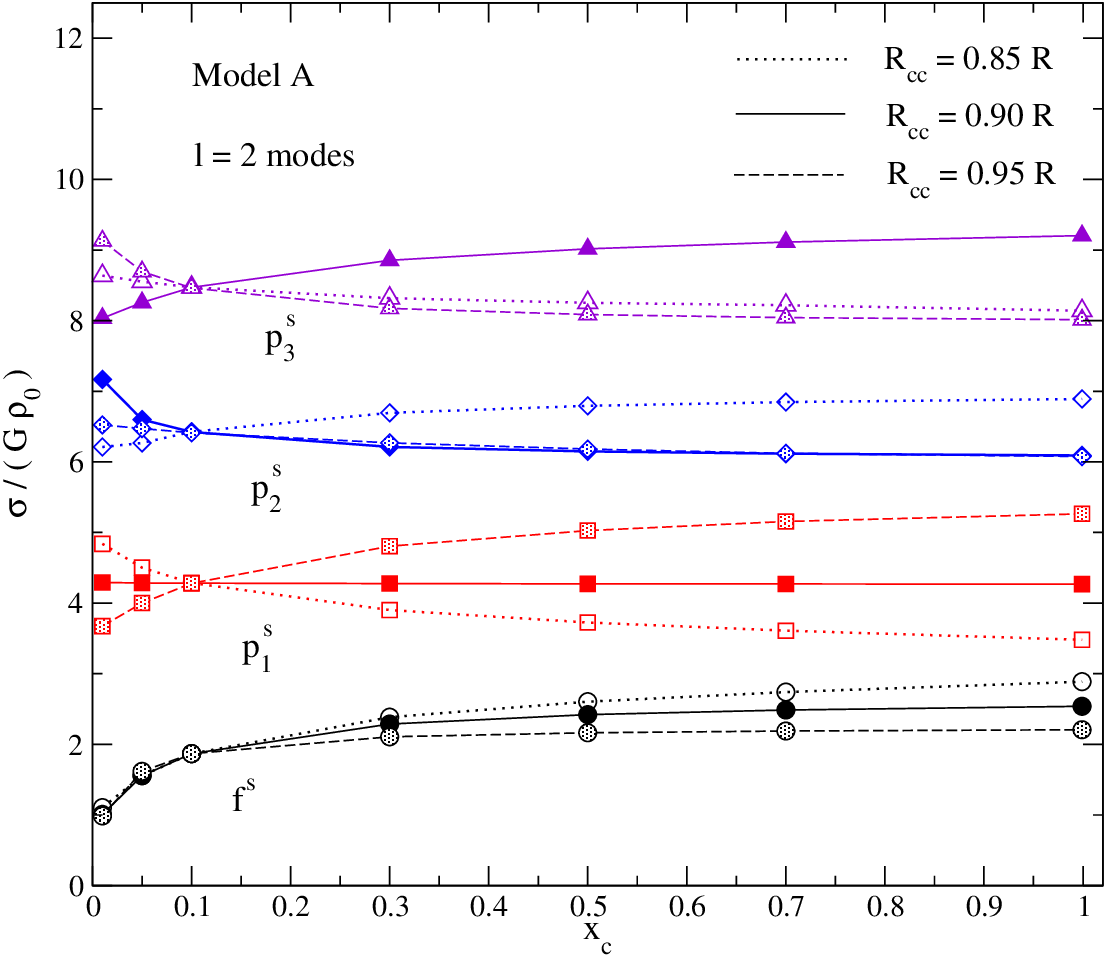}
\caption{Ordinary and superfluid $l=2$ acoustic modes for models A with
  zero entrainment, $x_{\p}=0.1$ and $\hat \mu_{\star} = 10^{-4}$. The
  left panel displays the dependence of the f- and p-mode frequencies
  on the crust composition $x_{\ch}$ for a model A with $R_{cc} = 0.9
  R$.  The right panel shows the superfluid $l=2$ f- and p-modes for
  models A with different $x_{\ch}$ and crust thickness,
  i.e. crust/core interface position $R_{cc}$ (see legend). The mode
  frequencies are shown in dimensionless units. For
  non-stratified models, the ordinary modes depend weakly on the crust
  composition (left panel), while the superfluid mode frequencies
  change significantly with respect to $x_{\ch}$ and $R_{cc}$.
\label{fig:7}}
\end{center}
\end{figure}

\subsection{Acoustic modes}

Superfluid two-constituent stars can sustain two classes of
fundamental and pressure modes, the ordinary and superfluid modes,
respectively. The ordinary modes, for which the two fluid components
tend to move in phase, have spectral properties similar to the f- and
p-modes of single fluid stars. Given previous results for that problem, we expect their
frequencies to be weakly affected by the presence of the elastic
crust~\citep{1988ApJ...325..725M}. The effects of the crust on the
superfluid modes, for which the fluid elements oscillate in
counter-phase, have not been studied previously. In this section, we provide the first analysis of this problem.
We begin by considering the acoustic spectrum of non-stratified stellar models and then
extend the analysis to stars with composition gradients.

For a non-stratified model A with standard parameters, i.e. zero
entrainment, $x_\p = 0.1$, $R_{cc} = 0.9 R$ and $\hat \mu_{\star} =
10^{-4}$, we determine the
$l=2$ f- and p-modes for different confined nucleon fractions in the
crust, $x_\ch$. Results for the ordinary and superfluid modes are
shown in the left panel of Fig.~\ref{fig:7} and in Table~\ref{tab:4}.
\begin{table}
\begin{center}
  \caption{\label{tab:4} Frequencies of the ordinary and superfluid
    $l=2$ acoustic modes for a selection of stellar models A and
    D. Frequencies are given in units of $\sigma / \sqrt{G \rho_0}$,
    where $G$ is the gravitational constant and $\rho_0$ represents
    the central mass density.  The first column labels the stellar
    model. The non-stratified models A have constant proton fraction,
    $x_\p = 0.1$ and the confined nucleon fraction $x_{\ch}$ is shown
    in the second column. The stratified models D have the same
    central proton fraction, $x_\p = 0.2$ at $r=0$, while the value of
    $x_{\ch}$ at $r=R_{cc}$ is given in the second column. All models have zero entrainment and $\hat \mu_{\star}
    = 10^{-4}$.  } 
\begin{tabular}{ c  c  c c c c c  c c c  }
  \hline
  Model &   $ x_{\ch} $  &  $ {}^2 \rm{f}^{o} $ & ${}^2 \rm{p}_{1}^{o}$  & ${}^2 \rm{p}_{2}^{o}$ & ${}^2 \rm{p}_{3}^{o}$ &  ${}^2 \rm{f}^{s}$ & ${}^2 \rm{p}_{1}^{s}$ & ${}^2  \rm{p}_{2}^{s}$   & ${}^2 \rm{p}_{3}^{s}$ \\
  \hline
  A  &  0.1   &  1.3848  &  3.9069  & 6.1122 & 8.1952 & 1.8687 & 4.2831  & 6.4227  & 8.4708\\
  A  &  0.5   &  1.3848  &  3.9071  & 6.1123 & 8.1949 & 2.4209 & 4.2727  & 6.1482  & 9.0163 \\
  A  &  1.0   &  1.3848  &  3.9073  & 6.1125 & 8.1948 & 2.5398 & 4.2689  & 6.0909  & 9.2041\\ 
  \\
 D  &  0.1   &   1.3085  &  3.5649  & 6.4524 &   8.4873  & 1.9319  & 4.3871 & 6.8852  & 9.2062 \\
 D  &  0.5   &   1.3462  &  4.3439  & 7.2824 &   8.1749  & 2.2839  & 3.4936 & 6.1328  & 9.1061 \\
 D  &  1.0   &   1.3392  &  3.6177  & 7.8464 & 10.2406  & 2.3946  & 5.2819 & 6.1476  & 8.6563 \\
\hline
\end{tabular}
\end{center}
\end{table}
\begin{figure}
\begin{center}
\includegraphics[height=70mm]{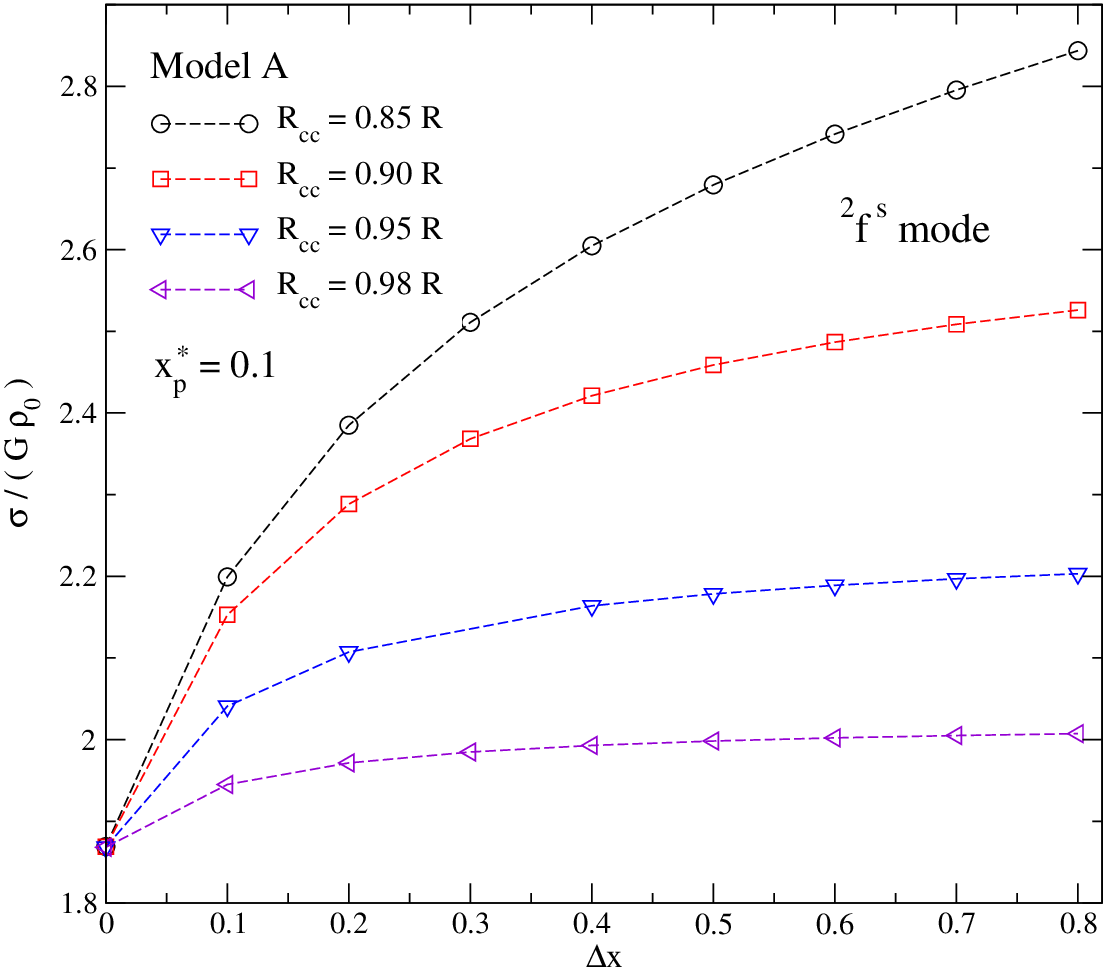}
\includegraphics[height=70mm]{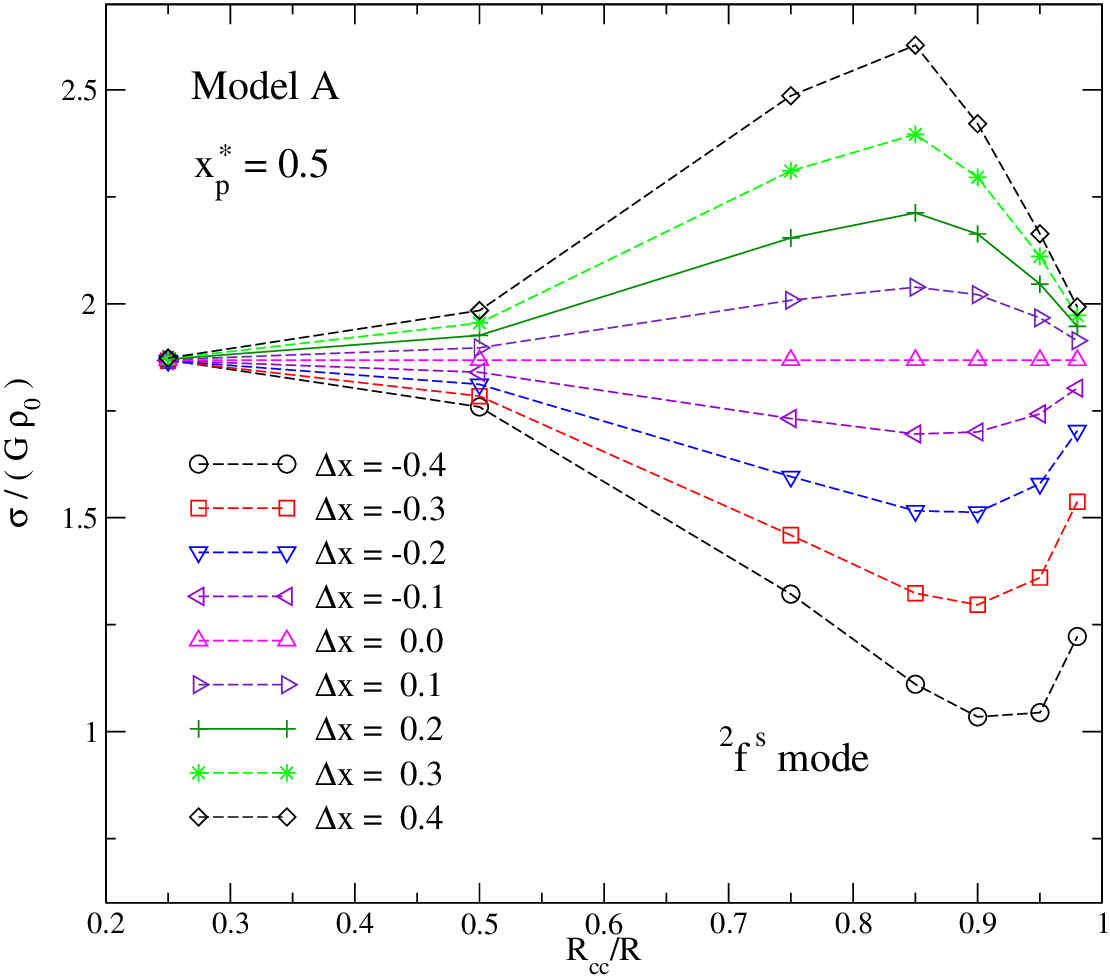}
\caption{ This figure displays the dependence of the superfluid $l=2$
  f-mode frequency on the crust/interface position $R_{cc}$ and on the
  composition step $\Delta x = x_{\ch} - x_{\p}$. The dimensionless
  eigenfrequencies are shown on the vertical axis. The background star
  is a model A with zero entrainment and $\hat \mu_{\star} =
  10^{-4}$. In the left panel, the $^2 \rm{f}^{s}$ mode frequencies
  are determined for two different sequences of models A and for four
  different crust thickness, $R_{cc}$ (see legend). The first sequence
  of stellar models has constant $x_{\p}\equiv x_{\p}^{\ast}=0.1$ and
  different $x_{\ch}$, while the second sequence has constant $x_{\ch}
  = 1-x_{\p}^{\ast} = 0.9$ and different $x_{\p}$. The quantity
  $x_{\p}^{\ast}$ is a parameter that identifies the two sequence of
  models A.  The mode frequencies for the first sequence are denoted
  with dashed-lines, while those for the second sequence with
  symbols. The right panel shows the effect of $\Delta x$ and $R_{cc}$
  on the $^2 \rm{f}^{\hspace{0.2mm} s}$ mode frequencies for two sets of models A
  with $x_{\p}^{\ast}=0.5$. This figure shows that for any two
  families of models A, denoted by $x_{\p}^{\ast}$, the
  eigenfrequencies are described by the same function of $\Delta$,
  i.e. $\sigma = \hat \sigma _{x_{\p}} \left( \Delta x \right) $.
\label{fig:8}}
\end{center}
\end{figure}
The dependence of the $ {}^2 \rm{f}^{o} $ and ${}^2 \textrm{p}_{n}^{\textrm{o}}$
modes on $x_\ch$ is very weak, as expected. For instance, the variation of the 
p-mode frequency  for  the two models, respectively, with   
$x_\ch = 0.1$ and $x_\ch=1.0$ is less than 0.01\%.
Focusing our attention
on the ordinary modes, we also study the effects of the crust
thickness on the eigenfrequencies. We consider a model A with $x_\p =
0.1$ and $x_\ch=0.2$, and change the interface location from
$R_{cc}=0.1 R$ to $R_{cc}=0.99 R$. The deviation away from
the ordinary mode frequency is very small even in the extreme cases, less than 0.05\%.

The behaviour of the superfluid modes is different.
Figure~\ref{fig:7} and Table~\ref{tab:4} show that the $ {}^2
\rm{f}^{\hspace{0.2mm} s} $ and $ {}^2 \textrm{p}_{n}^{\textrm{s}}$ modes depend significantly on
$x_\ch$, although in different ways. For instance, the frequencies of the
${}^2 \rm{f}^{\hspace{0.2mm} s} $ and $ {}^2 \textrm{p}_{3}^{\textrm{s}}$ $ \left(
  ^2\textrm{p}_{2}^{\textrm{s}} \right)$ modes are increasing (decreasing)
functions of $x_\ch$, while the frequency of the $ {}^2
\textrm{p}_{1}^{\textrm{s}}$ mode remains almost constant. The behaviour becomes even
more complex when we vary the crust thickness. In the
right panel of Fig.~\ref{fig:7}, we show the superfluid $l=2$ f- and
p-modes for three models A with different crust thickness,
i.e. $R_{cc} / R = 0.85, 0.90$ and $0.95$,
respectively. The crust region clearly affects the spectrum of
superfluid modes and it is not (yet) clear to us whether a simple relation
can be found to describe the observed behaviour. For the $ {}^2 \rm{f}^{s}
$ mode, we see that thicker crusts and larger fractions of confined nucei both tend to
increase the mode frequency.

In order to understand the $ {}^2 \rm{f}^{\hspace{0.2mm} s} $ mode better,
we study its dependence on the change in composition (due to the chosen chemical gauge) between the core and
crust, expressed in terms of $\Delta x \equiv x_{\ch} - x_\p$.
For a given $R_{cc}$, we construct two sets of models A. In the first
set, we choose $x_{\p}=0.1$ and vary $x_{\ch}$, while in the second
model we choose $x_{\ch} = 0.9$ and vary $x_{\p}$.  We find that for
both these classes of models, the $ {}^2 \rm{f}^{\hspace{0.2mm} s} $ mode
frequency is described by the same function of $\Delta x$. Results for this function, which we 
denote $\sigma = \hat \sigma \left( \Delta x \right)$, are shown in the left panel of Fig.~\ref{fig:8}. 

We find a similar behaviour if we extend this analysis to other
stellar models, e.g., a sequence of models A with constant
$x_{\p} \equiv x_{\p}^{\ast}$ and a second sequence with constant
$x_{\ch} = 1-x_{\p}^{\ast}$. The parameter $x_{\p}^{\ast}$ identifies
the two sets of models.
For a given $x_{\p}^{\ast}$, the $ {}^2 \rm{f}^{\hspace{0.2mm} s} $ mode frequency
has the same functional dependence on $\Delta x$.  Note that the
frequencies are different for distinct classes of stellar models,
i.e. depend also on the parameter $x_{\p}^{\ast}$,
\begin{equation}
\sigma = \hat \sigma _{x_{\p}} \left( \Delta x \right) \, .
\end{equation}
In the right panel of Fig.~\ref{fig:8}, we show the $ {}^2 \rm{f}^{\hspace{0.2mm} s}
$ mode spectrum for two families of models A with $x_{\p}^{\ast} =
0.5$. In this case, we vary the composition fraction and the crust
thickness extending the parameter space towards unrealistic
configurations ($R_{cc} = 0.1 R$) in order to highlight the qualitative behaviour.

\begin{figure}
\begin{center}
\includegraphics[height=70mm]{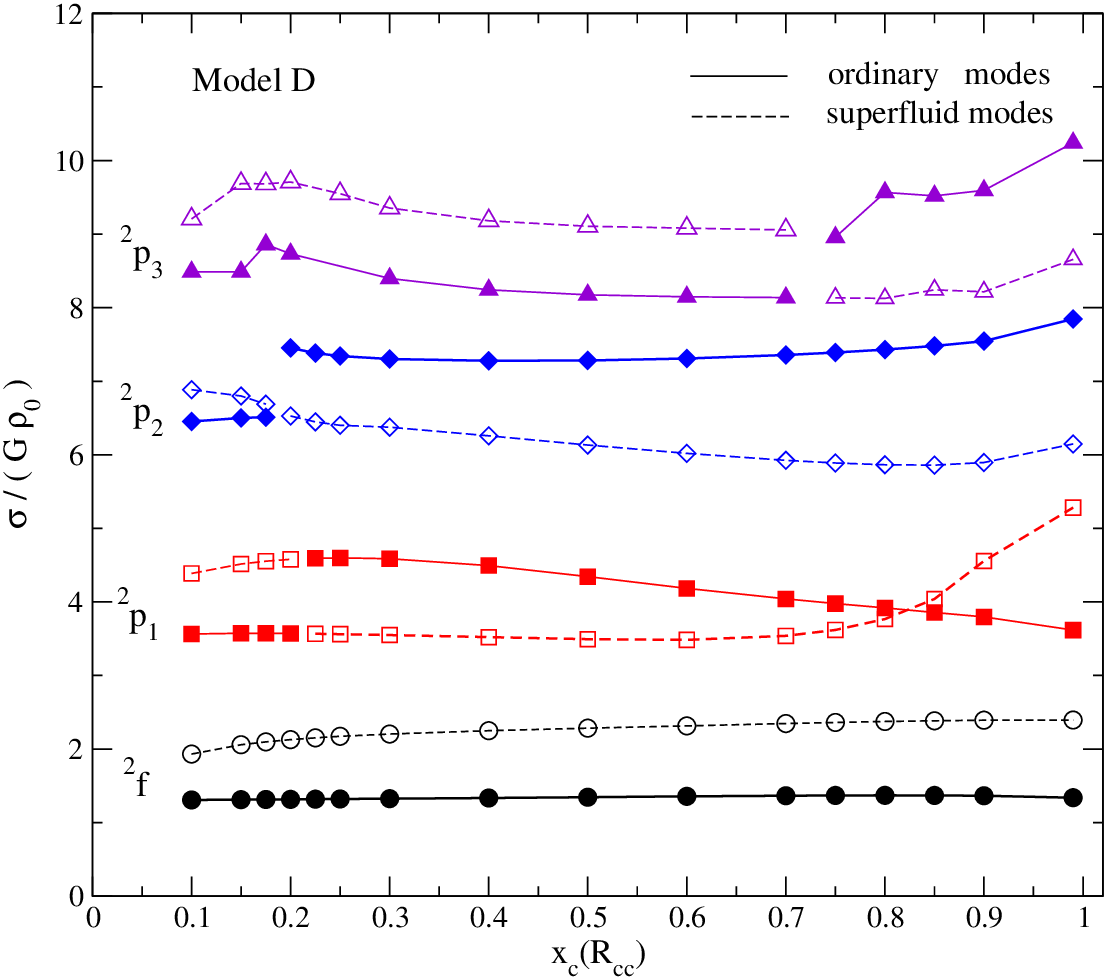}
\includegraphics[height=70mm]{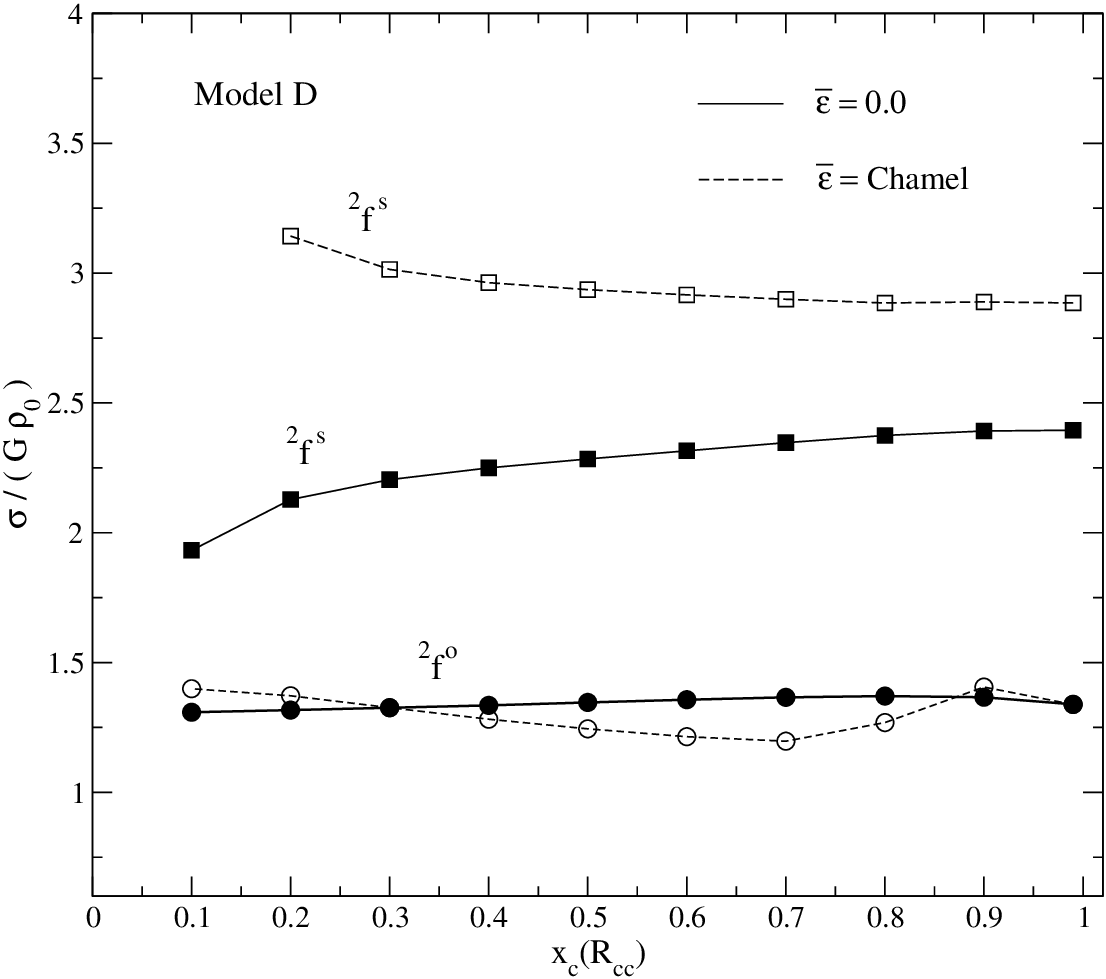}
\caption{ Frequencies of the ordinary and superfluid $l=2$ acoustic
  modes for stellar models with composition stratification. The left panel
  shows the mode frequency dependence on the crust component $x_{\ch}$
  determined at the crust/core interface, $r=R_{cc}$. On the vertical
  axis the frequencies are in dimensionless units. The background
  model D has zero entrainment, $\hat \mu_{\star} = 10^{-4}$ and
  $R_{cc} = 0.88 R$.  The p-mode spectrum exhibits avoided
  crossings between the superfluid and the ordinary modes.  These arise
  from the interaction between the co- and counter-moving degrees of
  freedom that are coupled in stellar models by composition
  gradients. The right panel shows the effect of a ``realistic''
  entrainment on the f-mode frequencies. The solid lines represent the
  results for zero entrainment models (the same as in the left panel),
  while the dashed lines denote the frequencies for models D with the
  Chamel entrainment configuration. The large entrainment of the deep
  regions of the crust affect both the superfluid and the ordinary
  $l=2$ f-mode. 
  We find that the variation of the 
  ${}^2\rm{f}^{\hspace{0.2mm}o}$ mode frequency  is about $12.4\%$ for a
  model D with $x_{\ch} = 0.7$, but decreases to
  $0.01\%$ for $x_{\ch} = 0.3$  \label{fig:9}}
\end{center}
\end{figure}

Let us now consider the acoustic mode spectrum of stratified stellar models. For
models D with zero entrainment, we construct a sequence of stellar
configurations with different composition $x_{\ch} $ at the crust/core
interface. The frequencies of the $l=2$ f-mode and the first three
p-modes for these models are shown in Fig.~\ref{fig:9} and Table~\ref{tab:4}.  The main
effect of composition stratification is the coupling between  the co- and
counter-moving degrees of freedom of the oscillations. The interaction
between ordinary and superfluid modes is evident in the spectrum
(see Fig.~\ref{fig:9}), where avoided crossings appear as the composition is varied. 
The chemical coupling also modifies
the frequencies of the ordinary modes, as seen from the data in
Table~\ref{tab:4}. For the $ {}^2 \rm{f}^{\hspace{0.2mm} o} $ mode, there is a
frequency variation of 2.8\% between models D with $x_{\ch} =
0.1$ and $x_{\ch} = 1.0$. However, this change may be due to the 
properties of the D models being somewhat different. In order to impose the continuity
conditions at the crust/core boundary, the total mass density of the
crust is a function of $x_{\ch}$ (see Table~\ref{tab:D}).
This mass variation of the crust can modify the $ {}^2 \rm{f}^{\hspace{0.2mm} o} $
mode, as its frequency is proportional to the mean density of the
stellar model,  $\sigma^2 \sim M/R^3$. We have not investigated this further, 
since the determined changes in the mode frequencies is not very significant. 

Now we turn to the effects that the large effective mass expected in the inner crust has on the acoustic spectrum.
We consider a set of models D with 
different composition $x_{\ch} $ at the crust/core interface and determine the frequencies of the 
ordinary and superfluid $l=2$ fundamental modes.  In the right panel of Fig.~\ref{fig:9} 
we compare the mode frequencies for the vanishing and realistic entrainment~\citep{2005NuPhA.747..109C} cases.  The results show that 
the effective mass has a noticeable impact on the frequency of the fundamental mode. For the 
 $ {}^2 \rm{f}^{\hspace{0.2mm} s} $ mode the frequency  changes significantly for the entire sequence of models D, while for the 
 $ {}^2 \rm{f}^{\hspace{0.2mm} o} $ mode we find a  variation of $12.4\%$ for a  model D with $x_{\ch} = 0.7$, but only $0.01\%$ for $x_{\ch} = 0.3$. 
This is an interesting result that deserves further investigation. However, in order to study more realistic neutron star models 
we need to work in the framework of  General Relativity.  
Developments in this direction are in progress, building on~\cite{2008PhRvD..78h3008L}.
  
Finally, we test the results against the acoustic spectrum obtained from time evolutions 
of superfluid neutron stars without crust~\citep{2009MNRAS.396..951P, 2011MNRAS.413...47P}. For the non-rotating model A, 
the frequencies of the ordinary and superfluid acoustic modes agree to better than $0.5\%$. This comparison is limited by the accuracy of the mode-extraction from the time-evolutions, so the result is acceptable.

\section{Conclusions\label{conclusions}} \label{sec:concl}

In this paper we have studied the effects of an elastic crust on the oscillation spectrum
 of superfluid neutron stars.  Building on recent results for the perturbations of two-fluid systems with 
 both elasticity and superfluidity~\citep{2011arXiv1105.1244A}, we considered Newtonian 
 models of mature neutron stars. The stellar core was modelled as a
  mixture of superfluid neutrons and a conglomerate of charged
  particles, while the inner crust was described by a lattice of nuclei
  permeated by a gas of free superfluid neutrons.  Focussing on nonrotating superfluid
  stars, we studied for the first time the effects of elasticity, entrainment and
  composition stratification on both shear modes and acoustic modes.  
 Our results show that the superfluid neutrons in the crust may have
  considerable impact on the star's spectrum.  The superfluid component clearly cannot be neglected if we want 
   to determine an accurate oscillation spectrum of mature neutron stars. If we compare 
   stellar models with and without superfluid neutrons in the crust, the difference between the shear mode frequencies 
   is very large, but becomes smaller (about 10\%)  when we consider a "realistic" effective neutron mass.    
   A superfluid component in the crust affects also the frequency of the superfluid acoustic modes, which are mainly dominated 
   by  a counter-phase motion between the two matter constituents. These superfluid modes are very sensitive to the composition 
   difference between the core and the crust, and to the crust thickness. Ordinary acoustic modes, 
   in which the stellar constituents oscillate mainly in phase, are less affected by the presence of the crust. 
   However, the spectrum of ordinary modes becomes more complex  when we consider stellar models 
   with composition stratification and more "realistic" entrainment. The spectrum exhibits avoided crossings between 
   superfluid and ordinary pressure modes due to the chemical interaction between  the co- and counter-moving degrees of freedom. 
   Furthermore, for some specific stellar models we have found that a large effective mass may affect  also
   the frequency of the ordinary f-mode at the 10\% level.

  The results presented in this paper are important as they represent the first analysis of the polar modes of 
  the coupled elastic superfluid system, complementing previous work on the corresponding axial mode problem.
  Our understanding of the latter problem remains more detailed, as it has also been considered in General Relativity. 
  This is important since one must work in the context of relativity in order to be able to consider realistic neutron star equations 
  of state in a meaningful way. The results we have presented were obtained for "relatively" simple, more phenomenological, stellar models. 
  These models have the advantage that one can explore the dependence on the key parameters, e.g., the composition, 
  but they are obviously not ``realistic''.  Nevertheless, the insights gained from the analysis provide useful guidelines 
  for developing and testing fully relativistic models in the future. The state-of-play as far as such developments are concerned is that the 
  required linear perturbation framework is complete and ready to be used in various applications of astrophysical interest~\citep*{And11-prep}. 
  This step is, obviously, non-trivial but we expect to make progress in the near future. In order to make these models truly realistic, there 
  also needs to be progress on various equation of state issues. We need more detailed models for the superfluid entrainment
  and the related effective masses, especially in the crust region.  
   Further work in this direction should certainly be  encouraged. After all,  the entrainment is one of the key parameters that 
   we may be able to  infer from future neutron star seismology.

\section*{Acknowledgements}
We are grateful to L. Samuelsson for fruitful discussions. 
AP was supported by the German Science Foundation 
(DFG) via SFB/TR7. NA acknowledges support from STFC through grant number PP/E001025/1.

\appendix

\section{Perturbation Equation}

In this section, we provide the complete linearised equations for the
polar sector.

\subsection{Core} \label{sec:Core}
In the core, the eigenvalue problem can be studied by the set of
perturbation variables defined in equation~(\ref{eq:ydef}), which we
re-state here for completeness:
\begin{eqnarray} 
 y_1  \equiv  \frac{W^{\p}}{r}  \qquad 
 y_2  \equiv  \frac{\delta \tilde \mu_\p}{r g}  \qquad
 y_3  \equiv  \frac{W^{\n}}{r}  \qquad 
 y_4  \equiv  \frac{\delta \tilde \mu _\n}{r g}  \qquad 
 y_5  \equiv  \frac{\delta \hat \Phi}{g r}  \qquad 
 y_6  \equiv  \frac{\delta \hat \Phi'}{g}  \,  .
\end{eqnarray} 
These dimensionless functions obey a system of six linearised equations:
\begin{eqnarray}
\frac{d  y_1}{dr} & = & 
                          - \left( \frac{\rho_\p'}{\rho_\p} + \frac{3}{r} \right)  y_1 
                          + g \left[  \frac{1-\veps_\n}{\hspace{-0.1cm}1-\bar \veps} \frac{l(l+1)}{r^2 \sigma^2} 
                                    - \frac{\mathcal{S}_{\p\p}}{\rho_\p} \right]  y_2
                           - g \left[  \frac{\veps_\p}{1-\bar \veps} \frac{l(l+1)}{r^2 \sigma^2} 
                                    + \frac{\mathcal{S}_{\p\n}}{\rho_\p} \right]  y_4 
			  + g \frac{l(l+1)}{r^2 \sigma^2}  y_5 \nn \\
\frac{d  y_2}{dr} & = & 
                           \frac{ 1-\veps_\p}{g} \sigma^2  y_1 
                          - \left( \frac{g'}{g} + \frac{1}{r} \right)  y_2 
                          + \frac{\veps_\p}{g} \sigma^2  y_3 
			  - \frac{ y_6}{r}  \\
\frac{d  y_3}{dr} & = &   - g \left[  \frac{\veps_\n}{1-\bar \veps} \frac{l(l+1)}{r^2 \sigma^2} 
                                    +  \frac{\mathcal{S}_{\n\p}}{\rho_\n} \right]  y_2  
                          - \left( \frac{\rho_\n'}{\rho_\n} + \frac{3}{r} \right)  y_3 
                          + g \left[  \frac{1-\veps_\p}{\hspace{-0.1cm}1-\bar \veps} \frac{l(l+1)}{r^2 \sigma^2} 
                                    - \frac{\mathcal{S}_{\n\n}}{\rho_\n} \right]  y_4 
			  + g \frac{l(l+1)}{r^2 \sigma^2}  y_5 \nn \\
\frac{d  y_4}{dr} & = &   \frac{\veps_\n}{g} \sigma^2  y_1 
                          + \frac{ 1-\veps_\n}{g} \sigma^2  y_3 
                          - \left( \frac{g'}{g} + \frac{1}{r} \right)  y_4 
			  - \frac{ y_6}{r}  \\
\frac{d  y_5}{dr} & = & - \left(  \frac{g'}{g} + \frac{1}{r} \right)  y_5 + \frac{ y_6}{r} \\
\frac{d  y_6}{dr} & = &   4 \pi r \left( \mathcal{S}_{\n\p} + \mathcal{S}_{\p\p} \right)  y_2 
                          + 4 \pi r \left( \mathcal{S}_{\n\n}  + \mathcal{S}_{\p\n}  \right)  y_4
                          + \frac{l ( l+1 )}{r}  y_5  -  \left(  \frac{g'}{g} + \frac{2}{r} \right)  y_6 
\end{eqnarray}

\subsection{Crust}  \label{sec:Crust}

In the crust, we use instead the dimensionless variables defined in
equation~(\ref{eq:zdef}), i.e., 
\begin{eqnarray} 
z_1 & \equiv & \frac{W^{\ch}}{r}  \qquad 
z_2  \equiv  - \frac{1}{q_\ch} \left( \rho_\ch \Delta \tilde \mu_\ch
  + \frac{2}{3} \mu  \nabla \cdot \xi_{\ch} - 2 \mu W'_{\ch} \right) \qquad 
z_3  \equiv  \frac{V^{\ch}}{r}  \qquad
z_4  \equiv \frac{\mu}{q_\ch} \left( V'_\ch - \frac{V_\ch}{r} + \frac{W_\ch}{r} \right)  \\
z_5 & \equiv & \frac{W^{\f}}{r}  \qquad 
z_6  \equiv  \frac{\Delta \tilde \mu _\f}{r g}  \qquad 
z_7  \equiv  \frac{\delta \Phi}{r g}  \qquad 
z_8  \equiv  \frac{\delta \Phi'}{g}  \,  ,  
\end{eqnarray} 
where in this work $q_{\ch} = p_{0}$. 

The system of equations that describe the oscillation modes of a
two fluid system in the crust is given by
\begin{eqnarray}
r \frac{dz_1}{dr} & = & - 3 \frac{\hat K_\ch}{\alpha_3} z_1 + \frac{z_2}{\alpha_3} +  l \left( l + 1 \right) \frac{\alpha_2}{\alpha_3}  z_3 
- r g \frac{\mathcal{S}_{\ch\f}}{\rho_\ch} \frac{\hat K_\ch}{\alpha_3} z_6 \\
r \frac{dz_2}{dr} & = &  \frac{\rho_\ch}{q_\ch}  \left[  - \left( 1-\veps_\ch \right) r^2 \sigma^2  + r^2 g' - 2 r g \frac{\alpha_2}{\alpha_3} 
                           + 4 \hat K_\ch \frac{\alpha_1}{\alpha_3} \frac{q_\ch}{\rho_\ch}  \left( 3 + \frac{d \ln \rho_\ch}{d \ln r}  \right) \right] z_1 \nn \\
                  & + & \left[  \frac{\rho_\ch}{q_\ch}  \frac{ r g}{\alpha_3} - \frac{d\ln q_\ch}{d \ln r } - 4 \frac{\alpha_1}{\alpha_3} 
                           + \left( 1 - \frac{4}{3} \frac{\alpha_1}{\alpha_3}  \right) \frac{d \ln \rho_\ch}{d \ln r} \right] z_2 
                    + \left[  r g \frac{\rho_\ch}{q_\ch} \frac{\alpha_2}{\alpha_3} 
                     - 2 \alpha_1 \left( 1 + \frac{1}{3} \frac{d \ln \rho_\ch}{d \ln r} \right) \left( 1 + 2 \frac{\alpha_2}{\alpha_3} \right)  
                              \right] l \left( l + 1 \right) z_3 \nn \\
                  & + & l \left( l + 1 \right) z_4 
                    - \frac{\rho_\ch}{q_\ch} \veps_\ch r^2 \sigma^2 z_5 
                    + \frac{ r g}{\alpha_3} \hat K_\ch \mathcal{S}_{\ch \f} \left[ \frac{4}{\rho_\ch} \frac{\alpha_1}{\alpha_3} 
                             \left( 1 + \frac{d \ln g}{d \ln r}  \right) - \frac{r g }{q_\ch} \right] z_6 
                    + g \frac{\rho_\ch}{q_\ch} r z_8  \\
r \frac{dz_3}{dr} & = & - z_1 + \frac{z_4}{\alpha_1} \\
r \frac{dz_4}{dr} & = & \left( r g \frac{\rho_\ch}{q_\ch} - 6 \hat K_\ch \frac{\alpha_1}{\alpha_3} \right)  z_1 
                    - \frac{\alpha_2}{\alpha_3}  z_2 
                    + \left[ -  \frac{\rho_\ch}{q_\ch} r^2 \sigma ^2 \frac{\hspace{-0.1cm}1 - \bar \veps}{1-\veps_\f} - 2 \alpha_1  
                             + 2 \alpha_1 \left( 1 + \frac{\alpha_2}{\alpha_3} \right)  l \left( l + 1 \right)  \right] z_3 
                    - \left( 3 + r \frac{q'_\ch}{q_\ch} \right) z_4 \nn \\ 
		  & - & r g   \frac{\rho_\ch}{q_\ch} \frac{\veps_\ch}{ 1 - \veps_\f} z_5
                    -  r g  \left( \frac{\veps_\ch}{ 1 - \veps_\f} \frac{\rho_\ch}{q_\ch} + 2 \hat K_\ch \frac{\mathcal{S}_{\ch\f}}{\rho_\ch} 
                                    \frac{\alpha_1}{\alpha_3}   \right) z_6
                    +  r g   \frac{\rho_\ch}{q_\ch} \frac{\hspace{-0.1cm}1 - \bar \veps}{1-\veps_\f} z_7  \\
r \frac{dz_5}{dr} & = & \frac{q_\ch}{\rho_\ch} \frac{\hat K_\ch }{\alpha_3} \frac{\mathcal{S}_{\f\ch}}{\rho_\f} 
                      \left[ 4 \alpha_1 z_1 + z_2 \right] 
                    -   \left( \frac{\veps_\f}{1-\veps_\f}  + 2  \frac{q_\ch}{\rho_\ch}\frac{ \mathcal{S}_{\f\ch}}{\rho_\f} \frac{\alpha_1}{\alpha_3} 
                   \hat K_\ch  \right) l ( l+1) z_3   
                    + \left[ - 3 + \frac{g}{1-\veps_\f} \frac{l(l+1)}{r \sigma^2} \right] z_5  \nn \\
                  &  + &g \left[ \frac{1}{1-\veps_\f} \frac{l(l+1)}{r \sigma^2} - r \frac{\mathcal{S}_{\f\f}}{\rho_\f}  +  \frac{4}{3} r \frac{q_\ch}{\rho_\ch}  
                            \frac{\alpha_1}{\alpha_3} \hat K_\ch \frac{ \mathcal{S}_{\f\ch}  \mathcal{S}_{\ch\f}}{\rho_\ch \rho_\f} \right] z_6 
                      + \frac{g}{1-\veps_\f} \frac{l(l+1)}{r \sigma^2} z_7 \\
r \frac{dz_6}{dr} & = & \left( \frac{\eps_\f}{g} r \sigma^2 - 4 \frac{q_\ch}{\rho_\ch}  \frac{\alpha_1}{\alpha_3} 
                              \hat K_\ch \frac{ \mathcal{S}_{\f\ch}}{\rho_\f} \right) z_1 
                    - \frac{q_\ch}{\rho_\ch}\frac{ \mathcal{S}_{\f\ch}}{\rho_\f} \frac{\hat K_\ch}{\alpha_3} z_2 
                   + \left( \frac{\veps_\f}{1-\veps_\f}  + 2  \frac{q_\ch}{\rho_\ch}\frac{ \mathcal{S}_{\f\ch}}{\rho_\f} \frac{\alpha_1}{\alpha_3} 
                   \hat K_\ch  \right) l ( l+1) z_3  \nn \\
                   & + & \left[  \left( 1 - \veps_\f \right) \frac{r \sigma^2}{g} + 2 - \frac{d \ln g}{d \ln r} 
                        - \frac{g}{1-\veps_\f} \frac{l(l+1)}{r \sigma^2}  \right] z_5  \nn \\
                   & - & \left[ \frac{g}{1-\veps_\f} \frac{l(l+1)}{r \sigma^2} + \frac{d \ln g}{d \ln r}  + 1 - r g \frac{\mathcal{S}_{\f\f} }{\rho_\f} 
                           + \frac{4}{3} \frac{q_\ch}{\rho_\ch} \frac{\alpha_1}{\alpha_3} \hat K_\ch \frac{ \mathcal{S}_{\f\ch}  \mathcal{S}_{\ch\f}}{\rho_\ch \rho_\f}
                        r g   \right] z_6
                   - \frac{g}{1-\veps_\f} \frac{l(l+1)}{r \sigma^2} z_7 - z_8 \nn \\
r \frac{dz_7}{dr} & = & - \left(  \frac{d \ln g}{d \ln r} + 1 \right) z_7 + z_8 \\
r \frac{dz_8}{dr} & = &   l ( l+1 )  z_7  - \left(  \frac{d \ln g}{d
    \ln r} + 2 \right) z_8  +  \frac{4 \pi r }{g} \delta \rho \label{eq:z8}
\end{eqnarray}
where the have defined the following quantities:
\begin{eqnarray}
\hat K_\ch \equiv \frac{\rho_\ch^2}{q_\ch} \mathcal{S}_{\ch\ch}^{-1} \, ,  \qquad \alpha_1 \equiv  \frac{\mu}{q_\ch} \, , \qquad \alpha_2 \equiv  \hat K_\ch - \frac{2}{3} \frac{\mu}{q_\ch} \, , \qquad 
\alpha_3 \equiv  \hat K_\ch + \frac{4}{3} \frac{\mu}{q_\ch}  \, . 
\end{eqnarray}
In equation~(\ref{eq:z8}), the perturbation of the total mass density
has the following form:
\begin{eqnarray}
\delta \rho & = & - \left[ r \rho'_{\ch} + 4 \left( \mathcal{S}_{\f\ch} +  \mathcal{S}_{\ch\ch} \right) 
               \frac{q_\ch}{\rho_\ch}  \frac{\alpha_1}{\alpha_3} \hat K_\ch  \right] z_1 
                 +  \frac{q_\ch}{\rho_\ch}  \frac{\hat K_\ch}{\alpha_3} \left( \mathcal{S}_{\f\ch} +  \mathcal{S}_{\ch\ch} \right)  
              \left[ - z_2 + 2 \alpha_1 \, l \left( l +1 \right) z_3 \right] 
                -  r \rho'_{\f} z_5  \nn \\
                & + & r g \left[ \left( \mathcal{S}_{\f\f}  + \mathcal{S}_{\ch\f} \right) 
              - \frac{4}{3} \left( \mathcal{S}_{\f\ch} +  \mathcal{S}_{\ch\ch} \right)  \frac{q_\ch}{\rho_\ch} 
                \frac{ \mathcal{S}_{\ch \f}}{\rho_\ch}
                \frac{\alpha_1}{\alpha_3} \hat K_\ch \right] z_6  \,  .
\end{eqnarray}

\nocite*

\label{lastpage}

\end{document}